\documentclass[a4paper,11pt]{article}
\usepackage[utf8]{inputenc}
\usepackage{lineno}
\usepackage[T1]{fontenc}
\usepackage{amsmath}
\usepackage{amsfonts}
\usepackage{amssymb}
\usepackage[]{graphicx} 
\usepackage{hyperref} 
\usepackage{authblk}
\usepackage{cases}
\usepackage{geometry}
\usepackage{subcaption}

\usepackage{nameref} 

\usepackage{lineno}

\let\oldequation\equation
\let\oldendequation\endequation
\renewenvironment{equation}
  {\linenomathNonumbers\oldequation}
  {\oldendequation\endlinenomath}

\usepackage[style=authoryear,natbib=true,sorting=nyt,url=true,giveninits=true,maxcitenames=2,maxbibnames=99,uniquename=init]{biblatex}

\title{Opposite variations for pore pressure on and off the fault during simulated earthquakes in the laboratory
}

\author[1]{Dong Liu}
\author[1]{Nicolas Brantut}
\author[2]{Franciscus M. Aben}
\affil[1]{Department of Earth Sciences, University College London, London, UK}
\affil[2]{TNO Applied Geosciences, Utrecht, 
 Netherlands}

\begin{document}

\maketitle

\begin{abstract} 
We measured the spatiotemporal evolution of pore pressure on- and off-fault during failure and slip in initially intact Westerly granite under triaxial conditions. The pore pressure perturbations in the fault zone and the surrounding bulk presented opposite signs upon shear failure, resulting in large pore pressure gradients over small distances (up to 10 MPa/cm). The on-fault pore pressure dropped due to localised fault dilation associated with fracture coalescence and fault slip, and the off-fault pore pressure increased due to bulk compaction resulting from the closure of dilatant microcracks mostly parallel to the maximum compression axis. Using a simplified analytical pore pressure diffusion model, we were able to capture our observations qualitatively. A quantitative fit could not be achieved, likely due to model simplifications and experimental variability. We show that a reduction in bulk porosity and relatively undrained conditions during failure are necessary for the presence of the off-fault pore pressure elevation. Considering this phenomenon as a consequence of a main shock, we further show that off-fault pore pressure increase have the potential to trigger neighbouring fault instabilities. In nature, we expect the phenomenon of off-fault pore pressure increase to be most relevant to misoriented faults, where the pre-rupture stresses can be large enough to reach the dilatancy threshold in the wall rocks. 
\end{abstract}

\subsection*{Plain Language Summary} 
The interplay between fluid flow and rock deformation plays a pivotal role in understanding the mechanics of geological processes and engineering activities, particularly in the context of seismic events. The slip of tectonic faults during earthquakes is expected to produce non-uniform variations of fluid pressure, notably along the fault plane and its vicinity. These variations may have a substantial impact on the dynamics of fault slip, an aspect that has thus far eluded precise quantification. We analyzed the evolution of pore pressure during simulated earthquakes in the laboratory, focusing on the contrast in pore pressure dynamics on and off the fault within a cm-sized sample of granite, a rock representative of the continental crust. Opposing trends were observed upon rock failure: The on-fault pore pressure dropped, which can be attributed to dilation within the fault zone (a well documented phenomenon) while the off-fault pressure increased, which was a surprise. The off-fault pressure increase can be explained by the rapid closure of small cracks upon unloading of the rock mass as the fault is formed. We confirm this interpretation using a simple model that adequately fits our data. During earthquakes in nature, we expect off-fault fluid pressure to increase in regions where initial stresses are elevated, for example where fault geometry is complex. 

\subsection*{Key points:}
\begin{itemize}
    \item Opposite variations of pore pressure occur on and off the fault upon shear failure.
    \item On-fault pore pressure drop results from fault dilatancy.
    \item Off-fault pore pressure elevation necessitates bulk porosity reduction and undrained conditions during failure.
\end{itemize}

\section{Introduction}

Pore fluids play an important role in natural \citep[see][and references therein]{LoBe2002} and induced seismic activities \citep{Ells2013} by varying the fault frictional strength via the effective normal stress. Such variations can occur throughout the earthquake cycles \citep{NuBo1972,Sibson1975,SlBl1992} and may lead to a strongly heterogeneous distribution of pore pressure in space. 

Multiple processes may influence the pore pressure variations: dilation \citep{BrMa1968,Mart1980,SaDC2009,Bran2020,AbBr2021} and compaction \citep{FaSa2018,Proctor2020} of the fault gouge, frictional heating and thermal pressurization \citep{Lach1980}, fluid injection \citep{Ells2013}, fault sealing \citep{Sibson1975},  and dehydration \citep{KoOl1995,OlKo1995}. These processes can be generalized as the changes in the pore volume \citep{LoBe2002} and the fluid mass. In this paper, we focus on the pore pressure variations resulting from pore volume changes during shear failure.

Dilation, the process of porosity increase during deformation, can be localised inside the fault zone or distributed in the bulk medium. A dilatant fault zone acts as a sink of pore fluids during seismic slip and leads to a near-fault region characterised by a negative pore pressure perturbation \citep{Bran2020,AbBr2021}. 
However, the pore pressure in a dilatant bulk medium surrounding a (prospective) fault may respond differently to fault formation or instability. 
Most compact rocks dilate with increasing differential stress when it goes beyond a stress level known as C' \citep{Brace1966}. This dilation is due to the growth and opening of microcracks that are mostly parallel to the maximum compression axis \citep[][to cite a few]{Brace1966,HoNe1985,NeOb1988}. Such formation of dilatant cracks result in the anisotropy in the Skempton's coefficients $B$ \citep{HaWa1995,Chen1997,LoBe2003,Wong2017,Elsi2023} describing the responses of the pore pressure variation to a sudden stress change. Such an anisotropic characteristic has been accounted for by previous studies \citep{CoccRice2002}, where they assumed that the pore pressure rises with an increasing stress ($B \geq 0$). However, recent experimental work in Westerly granite \citep{Elsi2021} reports the possibility of negative $B$ values in the direction of the maximum compression axis: the pore pressure may thus increase upon a drop in the differential stress. Such a behaviour results from a reduction of bulk porosity due to the elastic closure of anisotropic microcracks which are formed parallel to the maximum compression axis. As a result, an increase of the pore pressure off-fault and a decrease on-fault may coexist upon a stress drop during the shear failure or fault instability. The pore fluid pressure distribution thus results from the interplay between the porosity decrease inside the bulk and the porosity increase inside the fault zone. However, such pore pressure elevation upon shear failure is rarely reported or discussed for earthquakes in nature \citep{WaNu1984,MaWa2015} and has not been measured previously in the laboratory. The interplay between the on-fault dilation and off-fault compaction upon shear failure remains unclear.


In this paper, we aim to detect and quantify the possible coexistence of on-fault pressure drop and off-fault pressure increase. To do so, we carried out triaxial failure experiments on intact Westerly granite and measured the local pore pressures during failure. Miniature pore pressure sensors \citep{Bran2020,BrAb2021} were positioned on the main failure plane as well as at a distance away from the fault, in the bulk medium that experiences unloading during failure. From these measurements, we observed the off-fault pore pressure elevation upon shear failure and the on-fault pore pressure reduction. The pore pressure records as a function of distance, and the stress history recorded during failure are then used to validate analytical models that capture this phenomenon. The model yields a qualitative alignment with the observations, yet a quantitative misfit is noted, likely attributed to model simplifications and inherent experimental variability. We then expand the model to the more general case of an infinite medium. We also attempt to analyse the potential triggering of neighbouring faults and aftershocks, and discuss the possibilities to capture the same phenomenon in natural earthquakes.

\section{Experimental set-up and methods}

Four samples of intact Westerly granite were tested in the triaxial Rock Physics Ensemble in the Rock and Ice Physics Laboratory at University College London. They were all cored into cylinders of 40 mm in diameter with ends ground parallel to a length of 100 mm. Two samples were cut with notches to favour a prospective rupture plane while the others were tested as-is. 
The notches were cut on the opposite sides of the samples (WG17 and WG18) with a depth of 17 mm and an inclination of 30~$\deg$ to the cylinder axis. We then filled the notch space with Teflon spacers before jacketing the samples. We placed pore pressure transducers \citep{BrAb2021,Bran2020,AbBr2021} close to and away from the prospective fault zone, aiming to capture the heterogeneity of the pore pressure distribution (see Figure~\ref{fig:Results}(a) for one example of the disposition). These transducers were placed directly in contact with the rock through holes in the nitrile sleeves, with epoxy sealing the space between sleeve and transducer. 

The equipped samples were placed in the triaxial cell for pressurization. We raised the confining pressure by pumping silicone oil into the pressure vessel and measured its value at the vessel inlet with a precision of 0.01~MPa. Pore pressure was imposed at the sample ends through a servo-hydraulic intensifier which recorded fluid pressure and volume variation. The samples were initially dry and were saturated in situ by flushing distilled water from the upstream end to the vented downstream end. Full saturation was achieved while steady-state flow was established (as measured from the intensifier volume) and a visible amount of water was collected on the vented side of the sample.
We then imposed the same constant pore pressure at the upstream and downstream ends until no flow was detected in or out of the sample. We repeated such a procedure by varying the confining stress and the imposed pore pressure step-wise to calibrate the pore pressure transducers \citep[see][for more details]{Bran2020,BrAb2021}.

Axial deformation was imposed via a servo-hydraulic ram and piston, and the corresponding shortening was measured with external linear variable differential transducers (LVDTs), corrected for elastic deformation of the loading column. The load was measured with an external load cell and corrected for piston seal friction. Shear stress on the fault was obtained from data recorded during and after failure, including a shear stress correction for the area change as a function of fault slip and the presence of low-friction Teflon spacers. We refer to \cite{AbBr2021,TeLW2010} for more details on stress correction.

All tests were performed with an imposed pore pressure $p_f$ at both ends of the sample (except WG16: $p_f=40$~MPa was only imposed at the upstream end of the sample and no flow was allowed at the downstream end) and a constant confining stress $P_c$ with $P_c-p_f=40$~MPa (Table~\ref{tab:results}).
All mechanical data, including the outputs of fluid pressure transducers, were recorded at a nominal rate of 1~Hz.

We aimed to measure the spatio-temporal evolution of the pore pressure during dynamic and stable failure, and to see under which conditions one could observe potential off-fault pore pressure elevation upon stress drop. We loaded all samples with an axial shortening rate of $10^{-6}\mathrm{s}^{-1}$. To favour a stable failure in WG16 and WG21, we paused the loading process shortly before the samples reached their peak stress, and reloaded them at the same strain rate after the full reequilibration of the pore pressure. In addition, measures were taken to increase the chance of capturing the pore pressure increase inside the bulk during stress drop: (1) In WG16, we unloaded the sample before the shear failure from nearly-peak stress with a strain rate of $10^{-5}\mathrm{s}^{-1}$, aiming to observe an increase of the bulk pore pressure. (2) In WG21, we increased the loading rate to $10^{-5}\mathrm{s}^{-1}$ during the failure process (see more details in Section~\ref{sec:results}) intending to trigger a dynamic instability and associated off-fault pore pressure elevation.

Apart from controlling the probable failure type and instability through loading histories, we also wanted to control different amplitudes of stress drop upon failure. WG16 and WG17 were notched and were expected to present a lower peak load and a smaller stress drop, while WG18 and WG21 were initially intact and were thus expected to experience a higher peak load and a larger stress drop. 
After the formation of a fault inside a sample (in WG16 and WG21, it corresponds to the moment when the differential stress decreased more gradually with time and was around the frictional resistance, and in WG17 and WG18, the moment when the differential stress dropped to almost zero due to the dynamic failure), we stopped the piston displacement and continued to measure the pore pressure evolution until complete reequilibration of pore pressure and stress. We summarize these experimental configurations in Table~\ref{tab:results}.

\begin{table}
    \centering
    \begin{tabular}{c|cccc}
    \hline
         Rock & WG16 &  WG17 & WG18 &  WG21 \\
         \hline
         Notch & Yes & Yes &  No & No \\
         Confining pressure $P_c$ (MPa) & 100 & 100 & 100 & 100 \\
         Pore pressure $p_f$ (MPa) & 60 & 60 & 60 &  60\\
         Loading pause near the peak load & Yes & No & No & Yes\\
         General failure type & Stable & Dynamic & Dynamic & Stable\\
         Max. slip rate (mm s$^{-1}$)  & 0.026 & >0.778 & $>1.7\times10^4$ & >0.17 \\
         Max. pressure elevation (MPa) & - & 24.3 & 50.7 & 9.4 \\
         Max. pressure drop* (MPa) & 28.0 & 30.5 & 15.2 & 14.1 \\
         Shear stress drop* (MPa) & 96.0 & 158.6 & 194.0 & 28.0 \\
         Shear stress at instability$^{\#}$ (MPa) & 167.8 & 158.6 & 194.0 & 141.0\\
         \hline
    \end{tabular}
    \caption{Summary of shear failure experiments. * indicates the pressure or stress drop during the period of pore pressure elevation, and for WG16, it corresponds to the pressure or stress drop during the stable failure. \# indicates the peak shear stress of the stable failure in WG16. 
    }
    \label{tab:results}
\end{table}

\section{Results}\label{sec:results}

\begin{figure}
    \centering
    \begin{tabular}{cc}
         \includegraphics[height=0.45\linewidth]{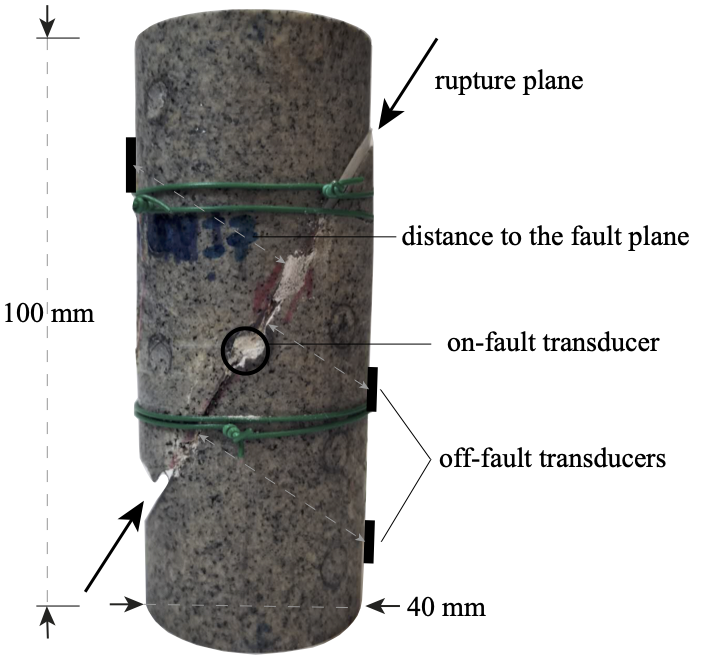} &
         \includegraphics[width=0.45\linewidth]{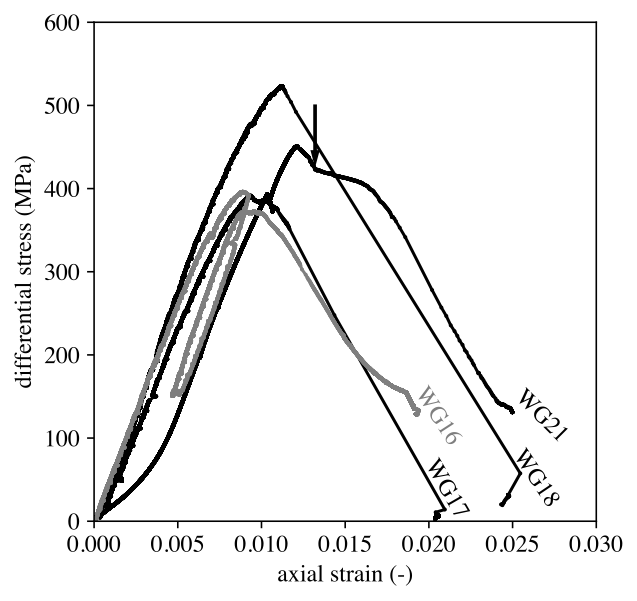}\\
        (a) & (b) 
    \end{tabular}
    \caption{(a) Illustration of the deformed sample of WG17 and its corresponding disposition of pore pressure transducers. (b) Stress versus strain curves for stable and dynamic shear failure experiments performed at 40~MPa effective confining pressure. Nominal strain rates were $10^{-6} \mathrm{s}^{-1}$ for all experiments except WG21, where the loading rate was increased to $10^{-5} \mathrm{s}^{-1}$ at the time indicated by the arrow to induce a dynamic instability.}
    \label{fig:Results}
\end{figure}

We report the stress evolution during the main shear failure in Figure~\ref{fig:Results}(b).  WG17 and WG18 failed dynamically and the differential stress dropped almost to zero after the shear rupture. WG16 and WG21 failed in a more stable manner with dynamic instability occurring during the failure process. Samples without the notches (WG18 and WG21) presented a significantly higher peak stress at failure, compared with those with the notches (WG16 and WG17). We summarize their different failure conditions in Table.~\ref{tab:results}.

\begin{figure}
    \centering
    \includegraphics[width=0.99
    \textwidth]{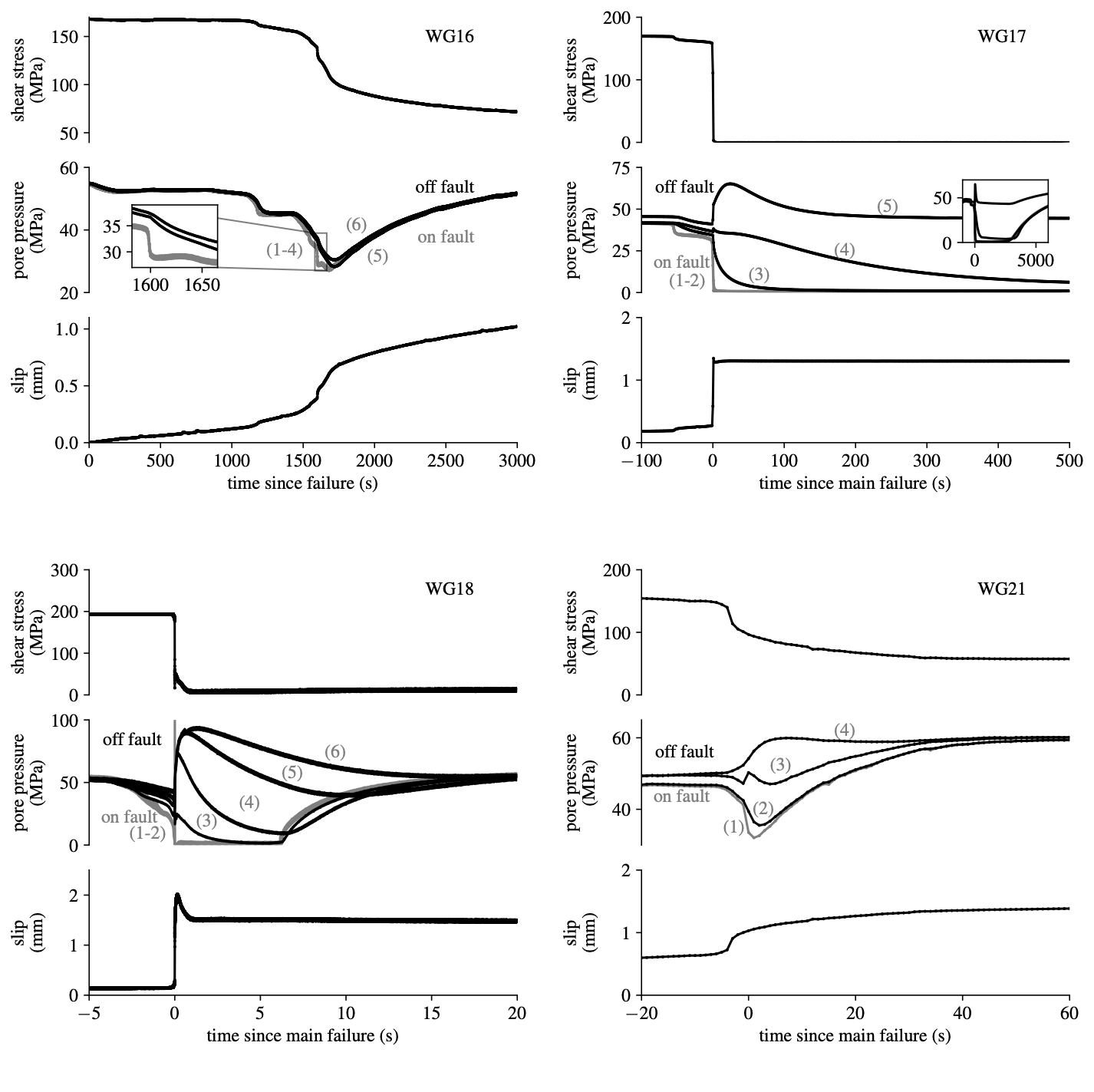}
    \caption{Records of the shear stress (top), pore pressure (middle), and fault slip (bottom) for stable  and dynamic failure at $P_{\mathrm{eff}}=40$~MPa. Middle panel: Off-fault pore pressure records shown as black curves, on-fault pore pressure shown as gray curves. Insets indicate a zoom-in or a zoom-out of the pore pressures around the main seismic event. The numbers in gray indicate the data measured by different pore pressure transducers, see the detailed locations of these transducers in Figure~\ref{fig:faultgeometry}.}
    \label{fig:PorePressureResults}
\end{figure}

We show in Figure~\ref{fig:PorePressureResults} the corresponding evolution of the pore pressure at different distances from the fault plane (detailed pore pressure locations and fault geometry are shown in Figure~\ref{fig:faultgeometry} and Table~\ref{tab:faultgeometry} of Appendix.~\ref{ax:geometrylocations}). The pore pressure evolution varies substantially between experiments (Table.~\ref{tab:results}), but we generally observe the same overall trend in all tests: pore pressure suddenly changes during failure, and returns to the homogenized distribution state after a sufficiently long time. In the following, we present each experiment separately, focusing on the different behaviours following the gradual pore pressure decrease. 

For WG16, we stopped loading the sample before the shear failure at a differential stress of $q=340$~MPa until full pore pressure equilibrium was reached across the sample. We then reloaded the sample to its peak stress at $q=396$~MPa and immediately unloaded it to $q=340$~MPa at a nominal strain rate of $10^{-5} \mathrm{s}^{-1}$. During this process, all pore pressure  records  present a similar evolution: the pore pressure drops to $\sim$53~MPa during the reloading and increases to $\sim$59~MPa during the faster unloading (Figure~\ref{fig:NoFailureWG16}(a)). After pore pressure reequilibriation, we unloaded the sample again with a strain rate of $10^{-5} \mathrm{s}^{-1}$ to a lower stress level $q=152$~MPa. This time, the pore pressure experienced a spatially uniform increase from $60$~MPa to $\sim 68$~MPa (Figure~\ref{fig:NoFailureWG16}(b)). After the full equilibration of the pore pressure with the intensifier, 
the sample was then reloaded to failure with a strain rate of $10^{-6} \mathrm{s}^{-1}$. All measured pore pressure gradually decreased with time before the peak stress was achieved. Note that some of the pore pressure transducers (Transducer number (2-4) as shown in Figure~\ref{fig:PorePressureResults}) were hydraulically connected by horizontal fractures formed during the failure process (see the failed sample in Figure~\ref{fig:faultgeometry}), which leads to a similar evolution of the on-fault and off-fault pore pressure.
The sample finally reached a peak stress and failed ($q=373$~MPa). The failure process was stable and the differential stress decreased gradually with time (over a duration of around 1~h). Around 0.5~h after peak load, a short instability occurred at $q=309$~MPa, the on-fault pore pressure experienced a sharp decrease of around 5~MPa compared to a drop in off-fault pore pressure of less than 1~MPa, as shown in the inset of Figure~\ref{fig:PorePressureResults}). We stopped loading the sample around 1~h after the peak load: The differential stress dropped slower than before and approached the fault frictional resistance ($q=155$~MPa).

\begin{figure}
    \centering
    \begin{tabular}{cc}
         \includegraphics[width=0.45\textwidth]{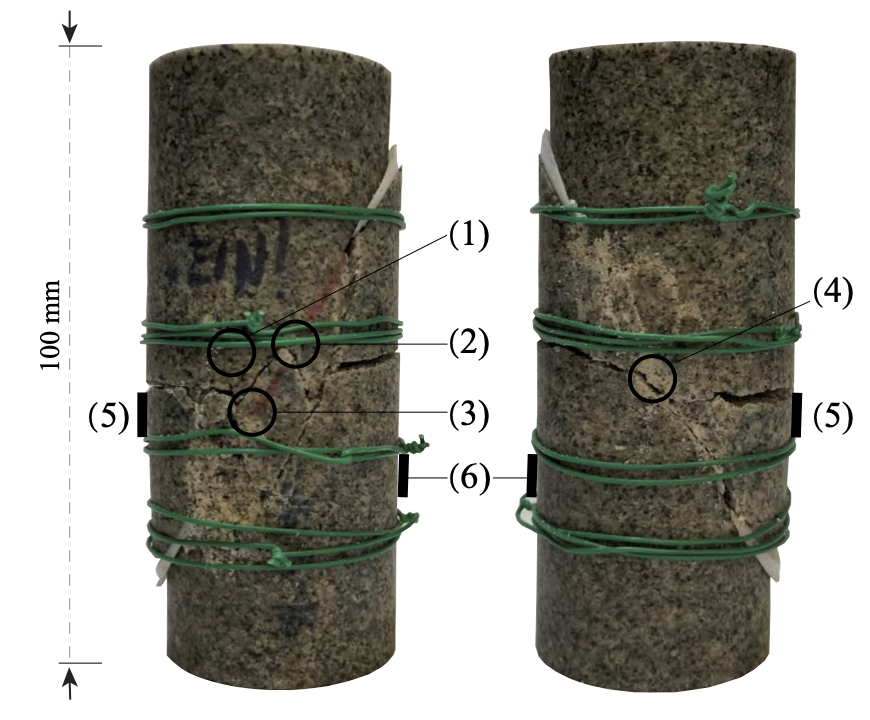} &  
         \includegraphics[width=0.45\textwidth]{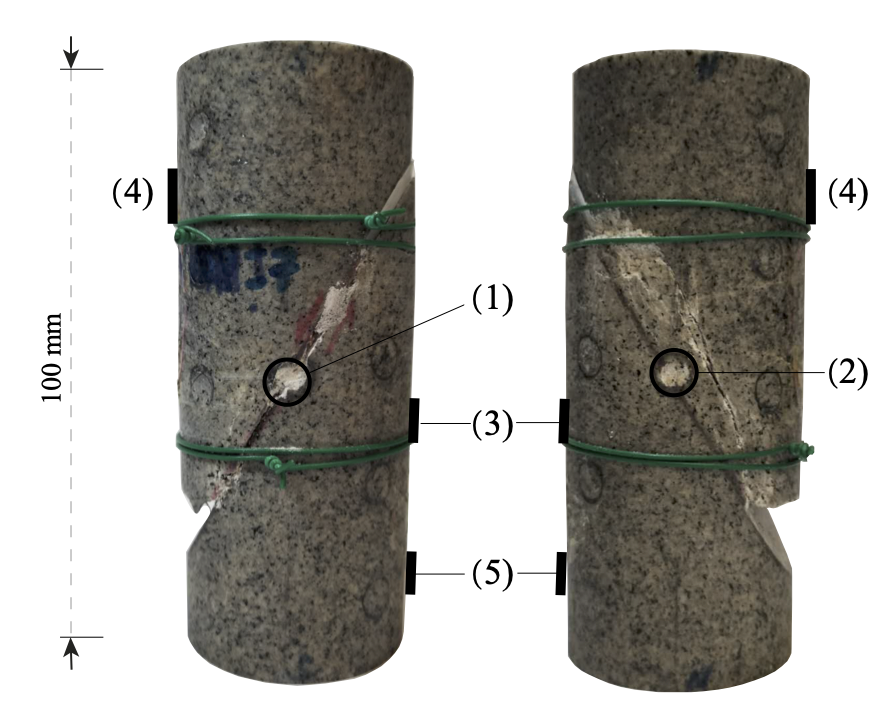} \\
         (a) WG16  & (b) WG17  \\
         \includegraphics[width=0.45\textwidth]{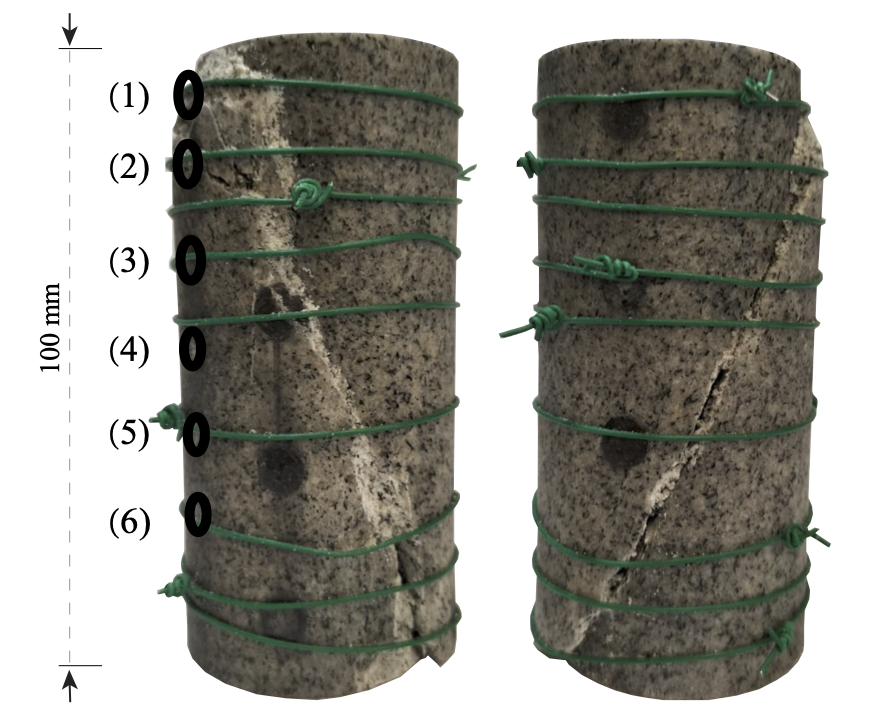} & 
         \includegraphics[width=0.45\textwidth]{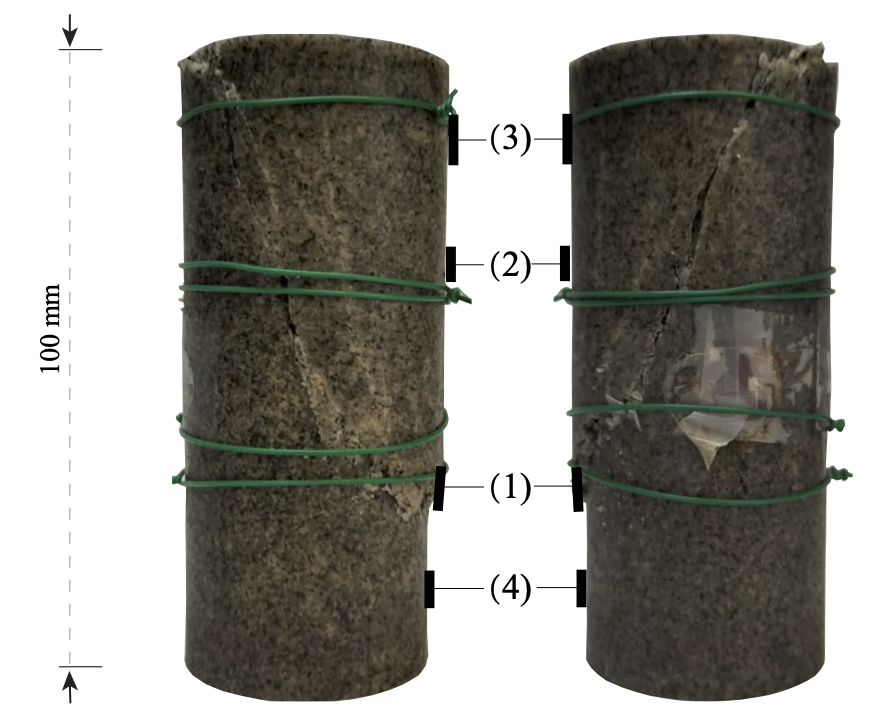} \\
         (c) WG18 & (d) WG21  \\
    \end{tabular}
    \caption{Fault geometry and locations of pore pressure transducers. The distances between the pore pressure sensors and the fault plane are shown in Table~\ref{tab:faultgeometry} of Appendix~\ref{ax:geometrylocations}.}
    \label{fig:faultgeometry}
\end{figure}

\begin{figure}
\centering
    \includegraphics[width=0.99\linewidth]{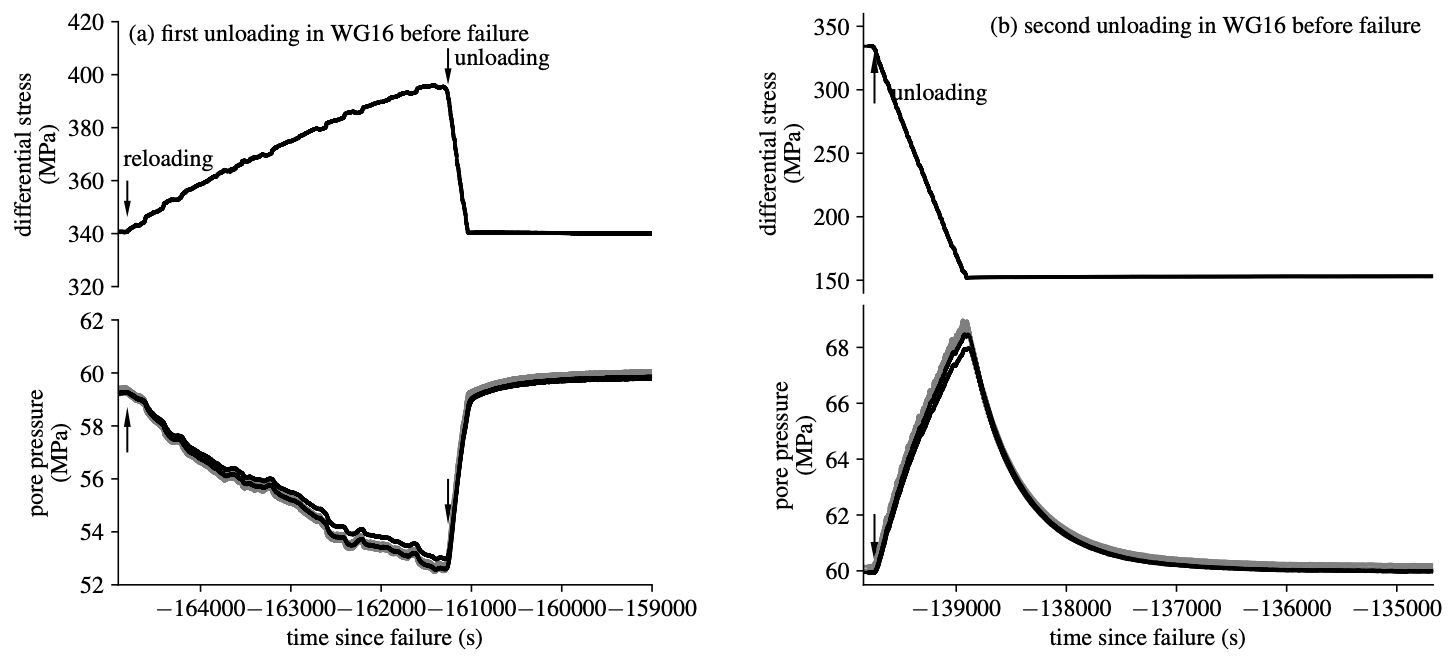} 
    \caption{Evolution of the pore pressure in the sample of WG16 upon unloading with a strain rate of $10^{-5} \mathrm{s}^{-1}$ before the shear failure. (a) illustrates the unloading from a stress level $q=396$~MPa very close to the peak load of the later shear failure $q=373$~MPa. (b) illustrates the unloading process following (a) after the equilibrium of the pore pressure distribution. The arrows indicate the time of unloading except that the one at $t=0$ in (a) refers to the start of reloading from $q=340$~MPa, at which we stopped loading the sample and waited for the establishment of an equilibrium pore pressure distribution. The gray and black curves in (a) and (b) correspond respectively to measurements of the same on-fault and off-fault pore pressure transducers as shown in Figure~\ref{fig:PorePressureResults}(a).} 
    \label{fig:NoFailureWG16}
\end{figure}

\begin{figure}
    \centering
    \includegraphics[height=0.45
    \textwidth]{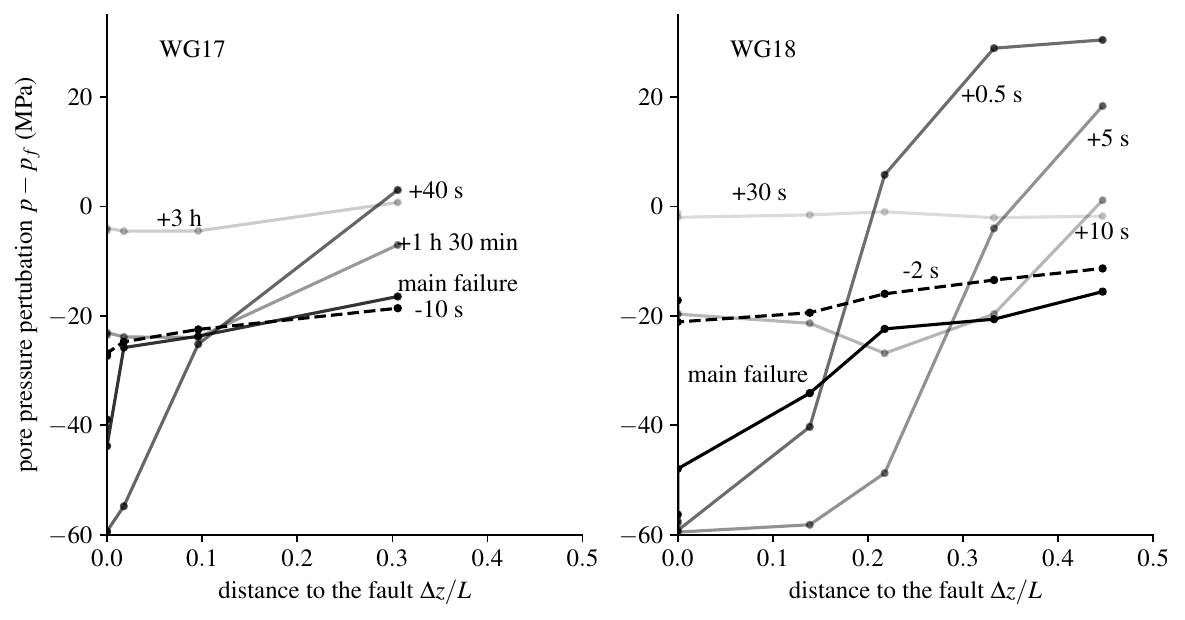}
    \caption{Time evolution of the pressure profile $(p-p_f)$ inside the sample. $\Delta z$ is the distance to the fault plane and $L\approx 50$~mm is the half of the sample length. The black dashed curve indicates the pressure distribution before the main failure. The black and grey solid curves indicate the pressure distribution at and after the main failure.}
    \label{fig:PorePressureProfile}
\end{figure}

Samples WG17 and WG18 were loaded at a rate of $10^{-6}\mathrm{s}^{-1}$. WG17 was notched to favour the formation of a planar fault and WG18 was originally intact. Sample WG18 experienced a higher peak stress ($q=523$~MPa) compared to WG17 ($q=391$~MPa).
Both samples failed dynamically with the dynamic failure occurring 21~min after the peak stress in WG17 at $q=248$~MPa and 2.5~min after the peak stress in WG18 at $q=504$~MPa.  Both WG17 and WG18 experienced a near-total stress drop, and we stopped deformation immediately after dynamic failure (Table.~\ref{tab:results}). 
The on-fault pore pressure dropped to around zero upon dynamic failure and led to the formation of a near-fault region characterised by a negative pore pressure variation (0.89-4.82~mm for WG17 and 6.93-10.86~mm for WG18, an approximate assessment based on the positions of the pore pressure sensors). Within this region, the pore pressure drop was smaller and occurred more gradually over time with increasing distance from the fault (Figure~\ref{fig:PorePressureResults}). In addition, transducers located outside this zone, further away from the fault, recorded a positive variation of the pore pressure. These pore pressure elevations were significant (24.3~MPa for WG17 and 50.7~MPa for WG18, Table.~\ref{tab:results}) and led to a pressure gradient of up to 10 MPa/cm inside the sample (Figure~\ref{fig:PorePressureProfile}). 
These pore pressure perturbations lasted for a much longer period than the failure process. After loading was stopped, the pore fluid pressure distribution reequilibrated gradually to its initial homogeneous state. However, the recovery period was quite different in these two samples: 15~s for WG18, and around half an hour for WG17. This difference in time probably resulted from a short-cut for fluid flow between the fault and the downstream sample end of WG18 which formed shortly after the shear failure (Figure~\ref{fig:faultgeometry}).

The loading of sample WG21 was paused at $q=372$~MPa before the sample reached its peak stress ($q=450$~MPa). It was then reloaded until failure after equilibrium of pore pressure was reached throughout the sample. A stable failure was first observed and all pore pressure records presented a gradual decrease with time. After 3~min of the gradual stress drop from the peak stress, we increased the loading rate to $10^{-5}\mathrm{s}^{-1}$ at $q=422$~MPa (see arrow in Figure~\ref{fig:Results}) to trigger a more significant stress drop within a shorter period. An instability occurred spontaneously around 5.5~min after the change of the loading rate, at $q=236$~MPa. Pore pressure elevation was observed in two off-fault transducers (transducers (3) and (4) in Figure~\ref{fig:faultgeometry}), but was short-lived and lasted only 10 seconds. A short-cut of fluid flow formed between the upstream end of the sample and the fault zone (Figure~\ref{fig:faultgeometry}), which led to the fast recovery of the pore pressure (around 40~s) to the homogenized state.

All tested samples presented a pore pressure drop on the fault upon shear failure, and the pore pressure distribution returned to the homogenized state after a sufficiently long time. Upon a dynamic instability, an opposite pore pressure variation on and off the fault was observed. Such observation was qualitatively reproducible in all samples that failed dynamically, but was quantitatively quite variable in different samples due to their different fracture geometries.

\section{Discussion}

\begin{figure}[!htp]
\centering
\includegraphics[width=\linewidth]{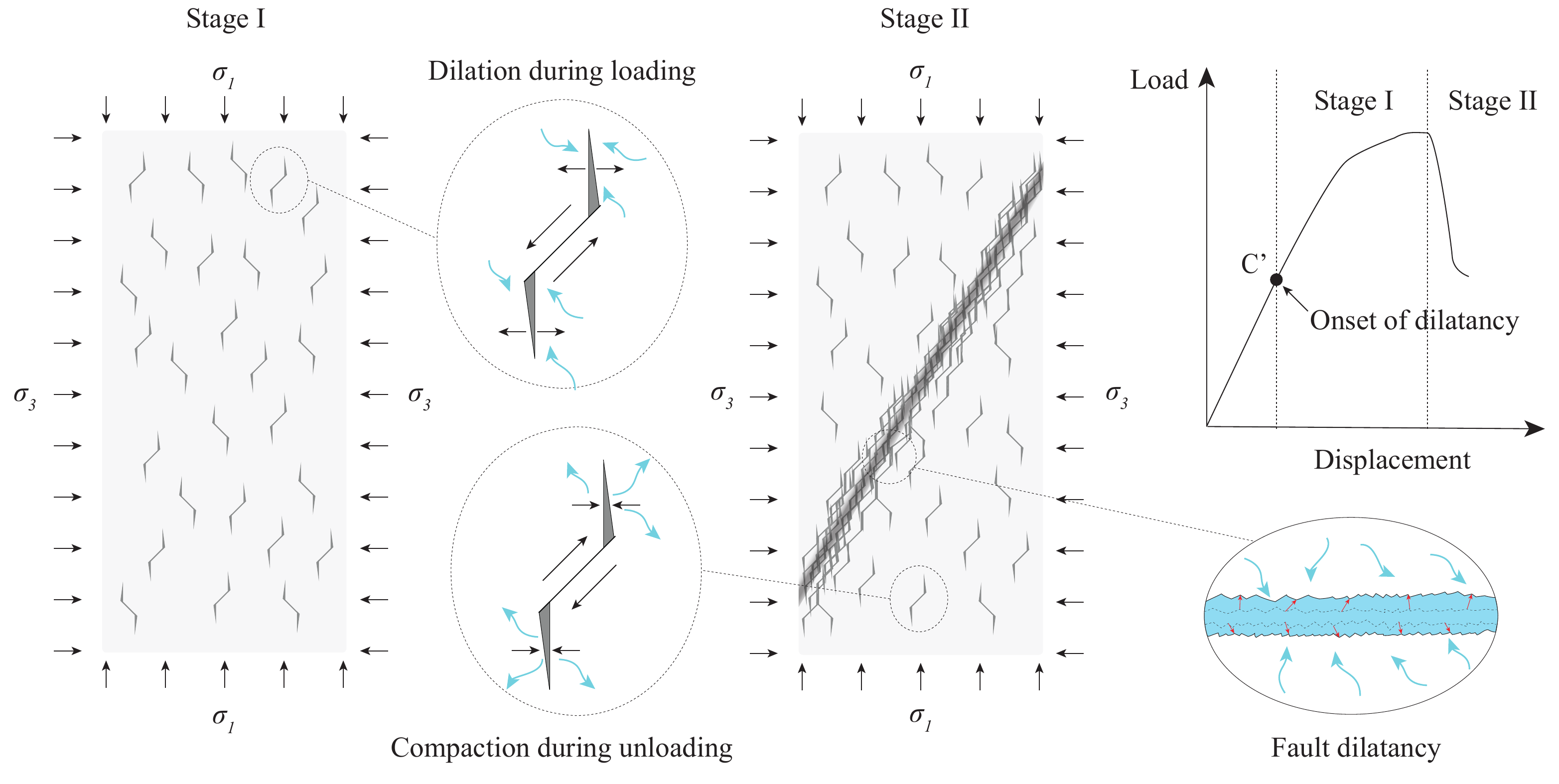} 
    \caption{Illustration of the mechanism of opposite variations for the pore pressure on and off the fault upon shear failure.} 
    \label{fig:Illustration}
\end{figure}

The occurrence of on-fault pore pressure drop during shear failure was consistently observed across all samples. This phenomenon is well-documented by previous studies \citep{Bran2020,Proctor2020,AbBr2021} and is attributed to the concurrent increase in porosity, a consequence of processes like fracture growth, fracture coalescence, and fault dilatancy \citep[e.g.][]{Brace1966} (Figure~\ref{fig:Illustration}).
Such on-fault pore pressure drop stabilizes fault slip \citep{RuCh1988,SeRi1995,Segall2010,FrZh17} and increases the fault frictional strength by increasing the effective normal stress on the fault plane \citep{Rice1975,Mart1980}. This effect, known as dilatancy hardening, is valid during the faulting process as long as the pore fluids remain relatively incompressible \citep{Bran2020}.  

In this study, we controlled the stable and dynamic failure of different samples by modulating the pore pressure at failure through loading histories. In WG17 and WG18, we imposed a continuous loading when the samples approached their peak stress. The pore fluid pressure kept decreasing during continuous loading due to bulk dilatancy and dramatically dropped to zero within the fault zone during failure. The relatively low pore pressure at the onset of rupture led to a limited dilatancy hardening effect \citep{AbBr2021} and a dynamic failure occurred. In WG16 and WG21, we modulated the pore pressure at failure to a higher value by pausing the loading at high pre-failure stress and waiting for the pore pressure to return to its equilibrium state, thereby removing the pore pressure reduction caused by bulk dilatancy. As a result, the dilatancy hardening effect was sustained for a longer time. In addition, in WG16, the damage zone formed during the loading towards the first peak load enhanced the fluid recharge and attenuated the pore pressure drop during the failure process. This further extended the period of dilatancy hardening effect and eventually led to a stable failure. 

To our knowledge, the occurrence of off-fault pore pressure elevation observed in WG17, WG18 and WG21 upon dynamic instability has not been reported in previous studies. This observation implies a reduction of porosity that probably comes from the closure of micro-cracks inside the bulk (Figure~\ref{fig:Illustration}). Mostly oriented parallel to the direction of the maximum compression \citep{PaWo2005,LoBe2003,Elsi2023}, these micro-cracks are generated prior to macroscopic sample failure when the stress level goes beyond the onset of volumetric dilatancy \citep{Bonn1974}. Consequently, a high stress level before the stress drop is necessary for the presence of the off-fault pore pressure elevation. Such an interpretation appears to explain the gradual pore pressure increase observed in WG16 during the imposed unloading stage prior to failure (Figure~\ref{fig:NoFailureWG16}): The sample experienced a porosity decrease upon a rapid unloading from a relatively high stress level. However, no off-fault pore pressure increase was observed in WG16 during the stress drop of its later failure process. This suggests that a high pre-failure/instability stress alone may not suffice for the presence of the coseismic off-fault pore pressure elevation.

Relatively undrained conditions during stress drop is another necessary component to observe the off-fault pore pressure elevation. It requires a small ratio between the shear failure duration and fluid diffusion time scale. Unlike the other experiments, in WG16, the stress dropped in a stable and gradual manner, making it possible for fluid diffusion to limit the pore pressure elevation associated with the gradual porosity reduction upon stress drop. 

We can now summarize the two necessary conditions for the presence of the coseismic off-fault pore pressure elevation as follows:
\begin{itemize}
    \item Porosity reduction upon stress drop from a relatively high level,
    \item Undrained conditions during failure or instability.
\end{itemize}
Note that the stress level before failure or instability should be at least as large as the threshold of dilatancy initiation. 

The fault zone thus acts as a localised sink of pore fluids upon shear failure, while the off-fault bulk volume acts as a distributed recharging source (Figure~\ref{fig:Illustration}). The pore pressure evolution then results from the interplay between fault zone dilatancy, off-fault porosity reduction, and fluid diffusion.

Based on this concept, in the following we analyse the key controls on such pore pressure variation in laboratory conditions by employing a simplified fluid diffusion model. We relate the pore pressure perturbations to the change of pore volumes in the sample. We compare the numerical simulation with the experimental data and discuss the possible reasons for misfit. We then extend the analysis to the case of an infinite medium, which may better describe the situation in nature. We also explore the possibility of such a mechanism to trigger the instability of neighbouring faults and aftershocks. We discuss the likely applied conditions of such a mechanism in nature and relevant observations during earthquakes.

\subsection{Models with 1-D fluid flow approximation}

In this section, we do not consider the 3D fault geometry and propose instead a simple 1-D fluid flow model to explain the data. We perform dimensional analysis to capture the governing dimensionless quantities to describe the interplay between the porosity increase in the fault zone, porosity reduction in the bulk, fluid diffusion, and applied boundary conditions. We also conduct numerical simulations to compare with the experimental results by taking the example of WG17, which is the sample that experienced failure in a simple geometry (Figure~\ref{fig:faultgeometry}). 

\subsubsection{Governing equations}
We consider a sample loaded under a constant confining stress $P_c$, a differential stress $q$, and imposed constant pore pressure $p_f$ at both ends.
We neglect the fault dip and assume a fault perpendicular to the cylinder axis located at the center of the sample. 
Accounting for the problem symmetry of the diffusion between the fault and the bulk, only half of the sample will be analysed with $z=0$ representing the fault position, and $z=L$ representing the upstream end of the sample.

The fluid flow will be simplified as the 1-D diffusion equation with the following initial and boundary conditions:
\begin{align}
    \frac{\partial p}{\partial t}-\frac{\kappa}{\eta S} \frac{\partial^2 p}{\partial z^2}&=\frac{\Delta \dot{\phi}_\mathrm{off}}{S}, \, z>0, \label{eq:fluidmass}\\
    \frac{\partial p|_{z=0}}{\partial t}-\frac{2 \kappa}{S_f w \eta}\left.\frac{\partial p}{\partial z}\right|_{z=0}&=-\frac{\Delta \dot{\phi}_\mathrm{on}}{S_f}, \label{eq:faultfluidmassa}\\
    p(z=L,t)&=p_f,
    \label{eq:faultfluidmassb}
\end{align}
where $\kappa$ is the host rock permeability, $\eta$ is the fluid viscosity and $w$ the width of the fault. $S$ and $S_f$ correspond to the storage capacity of the bulk and that of the fault. $w$ is the width of the fault. $\Delta \dot{\phi}_\mathrm{off}$ indicates the off-fault bulk porosity change rate, and $\Delta \dot{\phi}_{\mathrm{on}}$ the porosity change rate on the fault due to dilatancy.

The variation of differential stress, and the on-fault porosity change are assumed to be functions of the shear failure duration $t_\mathrm{fail}$.
\begin{equation}
\begin{aligned}
    \Delta q=\Delta q_\mathrm{max} f(t/t_\mathrm{fail}),\\
     \Delta \phi_\mathrm{on}=\Delta \phi_\mathrm{on}^\mathrm{max} h(t/t_\mathrm{fail}),
     \label{eq:shearfailurefunction}
\end{aligned}
\end{equation}
where $\Delta q_\mathrm{max}$ indicates the total stress drop during failure and $\Delta \phi_\mathrm{on}^\mathrm{max}$ indicates the maximum porosity variation inside the fault. $f(t/t_\mathrm{fail}), h(t/t_\mathrm{fail}) \in [0,1]$ are dimensionless functions describing the corresponding time evolution during the failure process.  

Here we start with a simple case to estimate the off-fault pore pressure elevation. We consider the case of an instantaneous porosity change in the bulk $\Delta \phi_{\mathrm{off}}$ upon shear failure:  The instantaneous pore pressure variation can be expressed as $\Delta \phi_\mathrm{off}/S$, where 
the porosity change $\Delta \phi_\mathrm{off}$ upon shear stress drop is a function of the effective confining stress and the differential stress drop. It can be estimated using the volumetric strain variation measured from cyclic loading tests in dry Westerly granite \citep{BrPe2023}. For an effective confining stress of 40~MPa, the porosity variation due to complete unloading from the peak differential stress $q_\mathrm{max}$ is around $0.86$\%  (see Appendix.~\ref{ax:porosity} for more details).
We would thus predict an instantaneous increase of the off-fault pore pressure of around 396~MPa assuming a differential stress drop of $5 q_\mathrm{max}/6$ and a bulk storage coefficient of $S=2\times 10^{-11}\, \mathrm{Pa}^{-1}$ \citep{AbBr2021}. However, the pore pressure increase observed during the experiments is much smaller than this simple prediction. One possibility is that the system is not perfectly undrained during the porosity change. Another possibility might be a porosity decrease smaller than that anticipated from dry cyclic loading data. A feedback may exist between the porosity variation and the local pore pressure, whereby the pore pressure increase originally resulting from the closure of microcracks tends to limit further microcrack closure and thus the overall reduction of porosity. In the following, we refine our predictions by allowing for a partially drained system in our models by simulating fluid diffusion inside the sample, and we refine the bulk porosity variation by accounting for a feedback between the pore volume change and the local pore pressure.

\subsubsection{Transient reduction of the bulk porosity}
The porosity reduction of the bulk increases with a larger stress drop and a larger effective confining pressure. It can be approximately expressed as a function of the differential stress change $\Delta q$ and the local pore pressure $p$ as follows when fitting with the experimental data \citep{BrPe2023} (see Appendix.~\ref{ax:porosity} for more details):
\begin{equation}
    \Delta \phi_\mathrm{off}=\Delta \phi_\mathrm{off}^\mathrm{mid} (1-\mathrm{e}^{(p-P_c)/P_o})(\Delta q/q_\mathrm{max})^\alpha,
    \label{eq:inelasticporosity}
\end{equation}
where $P_c$ is the applied confining stress and $\Delta q/q_\mathrm{max}$ is the dimensionless differential stress scaled by the peak differential stress $q_\mathrm{max}$ at the main failure. $\Delta \phi_\mathrm{off}^\mathrm{mid}$, $P_o$ and $\alpha$ are quantities related to material properties and experimental conditions: $\Delta \phi_\mathrm{off}^\mathrm{mid}$ is related to the maximum porosity reduction in the bulk. $\alpha$ describes the power-law dependency of the porosity variation on the stress drop. $P_o$ characterises the sensitivity of the porosity change to the local pore pressure, where a large value indicates little sensitivity to the change of local pore pressure $p$.
We then obtain the porosity change rate by differentiating Eq.~\eqref{eq:inelasticporosity} with time:
\begin{equation}
    \Delta \dot{\phi}_\mathrm{off}=\Delta \phi_\mathrm{off}^\mathrm{mid}(-\frac{\dot{p}}{P_o}\left(\frac{\Delta q}{q_\mathrm{max}}\right)^\alpha \mathrm{e}^{(p-P_c)/P_o}+\alpha \left(\frac{\Delta q}{q_\mathrm{max}}\right)^{\alpha-1} \frac{\Delta \dot{q}}{q_\mathrm{max}}(1-\mathrm{e}^{(p-P_c)/P_o})).
    \label{eq:porosityrate}
\end{equation}
We see that the bulk porosity variation does not present a step-like time evolution upon an instantaneous stress drop but becomes a transient process dependent on the local pore pressure.

\subsubsection{Dimensional analysis}

We define two pressure scales corresponding respectively to the undrained pore pressure decrease on the fault due to dilatancy $\Delta \Pi_\mathrm{on}^\mathrm{U}$ and the undrained pore pressure increase due to the porosity reduction inside the bulk $\Delta \Pi_\mathrm{off}^\mathrm{U}$. We also define the diffusion time scale $t_\mathrm{diff}$ for a finite sample size:
\begin{equation}
    \Delta \Pi_\mathrm{on}^\mathrm{U}=\frac{\Delta \phi_\mathrm{on}^\mathrm{max}}{S_f}, \, \Delta \Pi_\mathrm{off}^\mathrm{U}=\frac{\Delta \phi_\mathrm{off}^\mathrm{mid}}{S}, \, t_\mathrm{diff}=\frac{\eta S L^2}{\kappa} .
\end{equation}
We now scale different quantities with the following characteristic scales
\begin{equation}
    z \leftarrow z/L, \,  t \leftarrow t/t_\mathrm{diff}, \, p \leftarrow p/\Delta \Pi_\mathrm{on}^\mathrm{U}, \, \Delta \phi \leftarrow \Delta \phi_\mathrm{off}/\Delta \phi_\mathrm{off}^\mathrm{mid} ,
    \label{eq:scalequantities}
\end{equation}
and obtain seven dimensionless groups governing the fluid pressure distribution from equations (\ref{eq:fluidmass}-\ref{eq:inelasticporosity}):

\begin{equation}
    \frac{\Delta \Pi_\mathrm{off}^\mathrm{U}}{\Delta \Pi_\mathrm{on}^\mathrm{U}},\, 
    \frac{2 S L}{S_f w},\,  \frac{t_\mathrm{diff}}{t_\mathrm{fail}}, \, \frac{\Delta q_\mathrm{max}}{q_\mathrm{max}}, \,
    \frac{p_f}{\Delta \Pi_\mathrm{on}^\mathrm{U}}, \, \frac{P_c}{\Delta \Pi_\mathrm{on}^\mathrm{U}}, \,   \frac{P_o}{\Delta \Pi_\mathrm{on}^\mathrm{U}}.
    \label{eq:DmsGroups}
\end{equation}
$\Delta \Pi_\mathrm{off}^\mathrm{U}/\Delta \Pi_\mathrm{on}^\mathrm{U}$ characterises the amplitude of the potential pore pressure elevation off-fault. $2SL/S_f w$ characterises the localisation of pore pressure drop associated with fracture growth and fault dilatancy. $t_\mathrm{diff}/t_\mathrm{fail}$ characterises the period of the pore pressure perturbation. $\Delta q/q_\mathrm{max}$ represents the amplitude of the stress drop upon shear failure.
$p_f/\Delta \Pi_\mathrm{on}^\mathrm{U}$ and $P_c/\Delta \Pi_\mathrm{on}^\mathrm{U}$ characterise the magnitude of initial pore pressure $p_f$ and confining pressure $P_c$ which enter the model to ensure that (1) pore pressure $p$ does not become negative, and (2) to characterise the dependency of porosity on the effective confining pressure ($P_c-p$).
$P_o/\Delta \Pi_\mathrm{on}^\mathrm{U}$ is a material-dependent dimensionless property describing the sensitivity of the porosity change to local pore pressure and confining stress. Larger values of these dimensionless quantities (except for $p_f/\Delta \Pi_\mathrm{on}^\mathrm{U}$) favour the presence of a more significant off-fault pore pressure elevation. 

It is worth noticing that not all these dimensionless groups are independent of each other. For example, early vaporization of pore fluid ($p_f/\Delta \Pi_\mathrm{on}^\mathrm{U} \ll 1$) inside the fault zone may result in a dynamic failure characterised by a very short failure period ($t_\mathrm{diff}/t_\mathrm{fail} \gg 1$).  In addition, the maximum pore pressure elevation is not purely a function of 
$\Delta \Pi_\mathrm{off}^\mathrm{U}/\Delta \Pi_\mathrm{on}^\mathrm{U}$, but also a function of the shear failure period $t_\mathrm{diff}/t_\mathrm{fail}$, the stress drop amplitude $\Delta q_\mathrm{max}/q_\mathrm{max}$, and $(p-p_f)/P_o$ ($p/\Delta \Pi_\mathrm{on}^\mathrm{U}$, $p_f/\Delta \Pi_\mathrm{on}^\mathrm{U}$, and $P_o/\Delta \Pi_\mathrm{on}^\mathrm{U}$) describing the sensitivity of the material porosity variation to the change of pore pressure. The latter controlling dimensionless groups may have an even more significant impact on the pore pressure variation than $\Delta \Pi_\mathrm{off}^\mathrm{U}/\Delta \Pi_\mathrm{on}^\mathrm{U}$.

\subsubsection{Numerical results and comparison with experiments}
Among all three samples presenting pore pressure drop on the fault and pore pressure elevation off the fault upon dynamic shear failure, WG17 has the simplest fault geometry and boundary conditions of fluid flow: the fault is planar, and there is no short-cut of fluid flow connecting the fault zone and the sample ends. We thus compare the numerical results with the experimental data of WG17.

The 1-D fluid flow problem can be simplified as a group of ordinary differential equations with initial conditions given by the pore pressure distribution interpolated from the experimental measurement. We consider two cases in our numerical simulations: (1) an instantaneous evolution of the bulk porosity and the fault dilatancy during a dynamic failure, and (2) a transient porosity reduction in the bulk with the formation of the fault zone within a certain period where we assume a linear time evolution of the stress drop and the dilatancy during the shear failure ($f(x)=h(x)=x$ for $0\leq x\leq1$, and $f(x)=h(x)=1$ for $x>1$ in Eq.~\eqref{eq:shearfailurefunction}). We refer to Appendix.~\ref{ax:numerical} for detailed formulation and numerical discretization. 

We set different simulation parameters for these two cases: For (1), the dynamic failure is instantaneous, and the pore pressure variation can be simulated by solving ordinary differential equations \eqref{eq:fluidmass1}.  We set the initial condition using the pore pressure profile prior to the failure and accounted for an instantaneous pore pressure elevation inside the bulk $\Delta p_\mathrm{off}^\mathrm{U}=23.5$~MPa, and a pore pressure drop inside the fault zone $\Delta p_\mathrm{on}^\mathrm{U}=55$~MPa (Eq.~\eqref{eq:faultfluidmass1}). For (2), the governing ordinary differential equations turn to Eq.\eqref{eq:fluidmass}-\eqref{eq:porosityrate}. We adopt pressure scales of $\Delta \Pi_\mathrm{off}^\mathrm{U}=396$~MPa and $\Delta \Pi_\mathrm{on}^\mathrm{U}=55$~MPa corresponding to bulk compaction and fault dilation upon failure. We then assume a total stress drop $\Delta q_\mathrm{max}/q_\mathrm{max}=1$ and a failure period of $t_\mathrm{fail}=1/5000 t_\mathrm{diff}$. $P_o=500$~MPa characterises the dependence of the porosity change on the local pore pressure: a larger value of $P_o$ indicates that the porosity variation is less sensitive to the change of the local pore pressure. For both (1) and (2), we use the same fluid flow transport properties with $S=2\times 10^{-11}\, \mathrm{Pa}^{-1}$ and $S_f w=3.5 \times 10^{-12}\, \mathrm{Pa}^{-1}$ ($2SL/S_f w=0.571$) \citep{AbBr2021}.

\begin{figure}[htp]
\centering
    \includegraphics[width=0.45\linewidth]{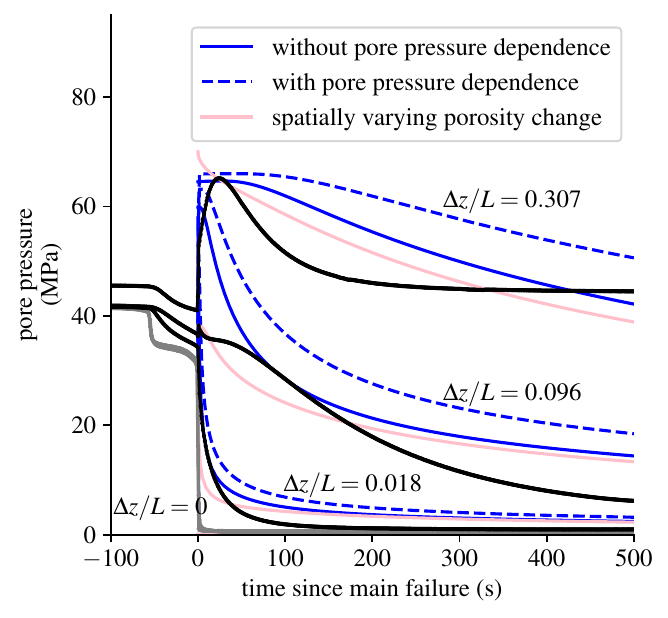}
    \caption{WG17: evolution of the pore pressure upon shear failure. The gray and black indicate respectively the on-fault and off-fault pore pressure measured in the experiments. The blue curves are obtained from the 1-D approximation of the fluid flow inside the sample with the solid and dashed curves representing respectively (1) an instantaneous failure and porosity variation, and (2) a more gradual stress drop and a transient porosity change. 
    The pink curves are numerical results accounting for (3) a spatially varying porosity change to simulate the effects of a fault damage zone while assuming an instantaneous failure and zero dependence of the porosity change on the pore pressure. $\Delta z$ is the distance to the fault plane and $L\approx 50$~mm is the half of the sample length.
    }
    \label{fig:DiffusionWG17}
\end{figure}

The simulated results reproduce an on-fault pore pressure drop and the off-fault pore pressure elevation upon shear failure, as shown in Figure~\ref{fig:DiffusionWG17}. They describe well the trend of the pore pressure evolution but not their quantitative variation. Following the instantaneous increase, the simulated off-fault fluid pressure decreases at a slower rate compared to the experimental observation. Various reasons may contribute to such a discrepancy.

(1) The stress drop during the shear failure occurs in a much shorter period than that in the cyclic loading tests \citep{BrPe2023}, from which we estimated the porosity reduction inside the bulk. In the case of a sliding crack model \citep{HoNe1985,NeOb1988,AsSa1990,BaGr1998,SiGS2001,DeEv2008,BhSR2011,LiBr2023}, the porosity variation at high stress comes from the opening or closure of microcracks parallel to the direction of maximum compression associated with the relative sliding of the preexisting defects. However, in the case of an instantaneous shear stress drop \citep{Kach1982a,DaBr2012,DaBr2020},  back-sliding may be limited along the preexisting defects' surfaces. It may lead to much less wing crack closure and porosity variation inside the bulk than what was measured in dry Westerly granite at a much smaller strain rate \citep{BrPe2023}. 

(2) There is a time delay for pressure sensors to respond upon a pore pressure change inside the sample. This delay is due to the flow of pore fluids from the sample to the sensor cavity, and depends on the permeability and diffusivity of the rock. For intact Westerly granite at an effective confining stress of 40~MPa, the response time is around 170~s \citep{Elsi2021}; here, the sample is damaged due to the loading process, so that we expect a faster response time, but the delay might be of the order of seconds to tens of seconds, which may damp the measured local pore pressure elevation upon instantaneous failure. 

(3) The over-simplification of the 3-D fluid flow by 1-D diffusion and the complexity and width of the fault zone are also responsible for the misfit between the numerical results and experimental data. 

(4) The possible presence of damage zone \citep{Bran2020,MiFa2009} around the main fault may increase the porosity off-fault and serves as a balancing term to the porosity reduction due to the shear stress drop. In the following, we discuss the effects of the damage zone in more details.

\subsubsection{Effects of the off-fault damage zone}

The heterogeneity of the porosity variation due to a damage zone may have a significant influence on the pore pressure distribution. To account for such a heterogeneity in porosity variation due to the presence of the damage zone, we perform numerical simulations by assuming an increasing porosity reduction (less damage) with increasing distance away from the fault zone upon dynamic failure. We neglect the dependency of porosity variation on the local pore pressure and assume an instantaneous porosity change upon failure. The pore pressure evolution can be thus approximated by solving the ordinary differential equations \eqref{eq:fluidmass1}. We use the same simulation parameters as the case of a uniform bulk porosity variation (shown in the blue solid curves of Figure~\ref{fig:DiffusionWG17}) except that the initial conditions are different. The initial pore pressure distribution then becomes the sum of the interpolated pre-failure pressure profile and their corresponding instantaneous changes upon failure: $\Delta p(\Delta z/L=0.0)=-30.23$~MPa, $\Delta p(\Delta z/L=0.018)=-8.59$~MPa, $\Delta p(\Delta z/L=0.096)=1.86$~MPa, $\Delta p(\Delta z/L=0.307)=28.99$~MPa, and $\Delta p(\Delta z/L=1.0)=0$~MPa indicating the imposed constant pore pressure at the sample ends. The numerical results accounting for the off-fault damage effect are shown with pink curves in Figure~\ref{fig:DiffusionWG17}. 

The simulated results with the non-uniform porosity variation seem to capture well the pore pressure evolution trend, but, as the other two configurations, they fail to reproduce the pore pressure variations quantitatively during the fluid recharging process. It is probable that a better fit of the experimental results calls for a more thorough numerical simulation accounting for the detailed 3D fault geometry and the bulk properties' evolution as a function of the damage zone.

\subsection{Extension to an infinite medium}

In the laboratory, the pore pressure response after shear failure  in a finite sample can be also viewed as a result of the interplay between different length scales of $L_\mathrm{fail}$, $L_\mathrm{dilat}$, and the half sample length $L$: 
\begin{equation}
    L_\mathrm{fail}=\sqrt{\frac{\kappa t_\mathrm{fail}}{\eta S}}, \, L_\mathrm{dilat}=\frac{S_f w}{2 S},
\end{equation}
where $L_\mathrm{fail}$ characterises the distance away from the ruptured fault which can be impacted due to the dilatancy during the shear failure period and $L_\mathrm{dilat}$ the maximum distance away from the ruptured fault that will be impacted by fault dilatancy. They are obtained by respectively setting $t_\mathrm{diff}/t_\mathrm{fail}=1$ and $2SL/S_fw=1$ in Eq.~\eqref{eq:DmsGroups}. 

In our experiments,  it is clear that $L_\mathrm{fail}\ll L_\mathrm{dilat}$, $L_\mathrm{fail}\ll L$ for dynamic failure. As a result, the pore pressure distribution inside the sample depends more on the interplay between the length scale $L_\mathrm{dilat}$ and the sample size $L$.
For typical values of Westerly granite and other compact crystalline rocks, $L_\mathrm{dilat}$ is often much larger than the sample size in the laboratory $L$. Sample size effects can be important in some cases (We refer to Appendix.~\ref{ax:boundary} for the detailed discussion of sample size effects) and makes it difficult to upscale directly the laboratory observation and measurement to the field scale ($L\to \infty$). It is thus necessary to extend the discussion to the case of an infinite medium. In the following, we focus on the assumption of an instantaneous variation of the porosity, and we discuss the influence of a transient porosity variation in Appendix.~\ref{ax:boundary}.

In the limit of an infinite medium ($L\to \infty$), an analytical solution exists for the off-fault pore pressure evolution upon a dynamic shear failure $t_\mathrm{fail} \to 0$. The perturbation of  pore pressure $\Delta p=p-p_f$ presents a timescale of $t_\mathrm{recharge}$, an influential length scale of $L_\mathrm{dilat}$, and a relative amplitude of $\Delta \Pi_\mathrm{off}^\mathrm{U}/\Delta \Pi_\mathrm{on}^\mathrm{U}$, which can be expressed as
\begin{equation}\label{eq:InfiniteAnalytical}
    \frac{\Delta p(z,t)}{\Delta \Pi_\mathrm{on}^\mathrm{U}}=-\left(\frac{\Delta \Pi_\mathrm{off}^\mathrm{U}}{\Delta \Pi_\mathrm{on}^\mathrm{U}}+1\right)\mathrm{e}^{z/L_\mathrm{dilat}+t/t_\mathrm{recharge}}\mathrm{erfc}\left(\frac{1}{2}\frac{z}{L_\mathrm{dilat}}\sqrt{\frac{t_\mathrm{recharge}}{t}}+\sqrt{\frac{t}{t_\mathrm{recharge}}}\right)+\frac{\Delta \Pi_\mathrm{off}^\mathrm{U}}{\Delta \Pi_\mathrm{on}^\mathrm{U}},
\end{equation}
where
\begin{equation}
    t_\mathrm{recharge}=t_\mathrm{diff}/\left(\frac{2SL}{S_f w}\right)^2=\frac{\eta S_f^2w^2}{4 S \kappa}
\end{equation}
is the recharge time scale, which can be expressed as a function of the diffusion time scale $t_\mathrm{diff}$ and the dimensionless storage ratio $2SL/S_f w$ for a finite size sample. 

\begin{figure}
\centering
      \includegraphics[width=0.95 \textwidth]{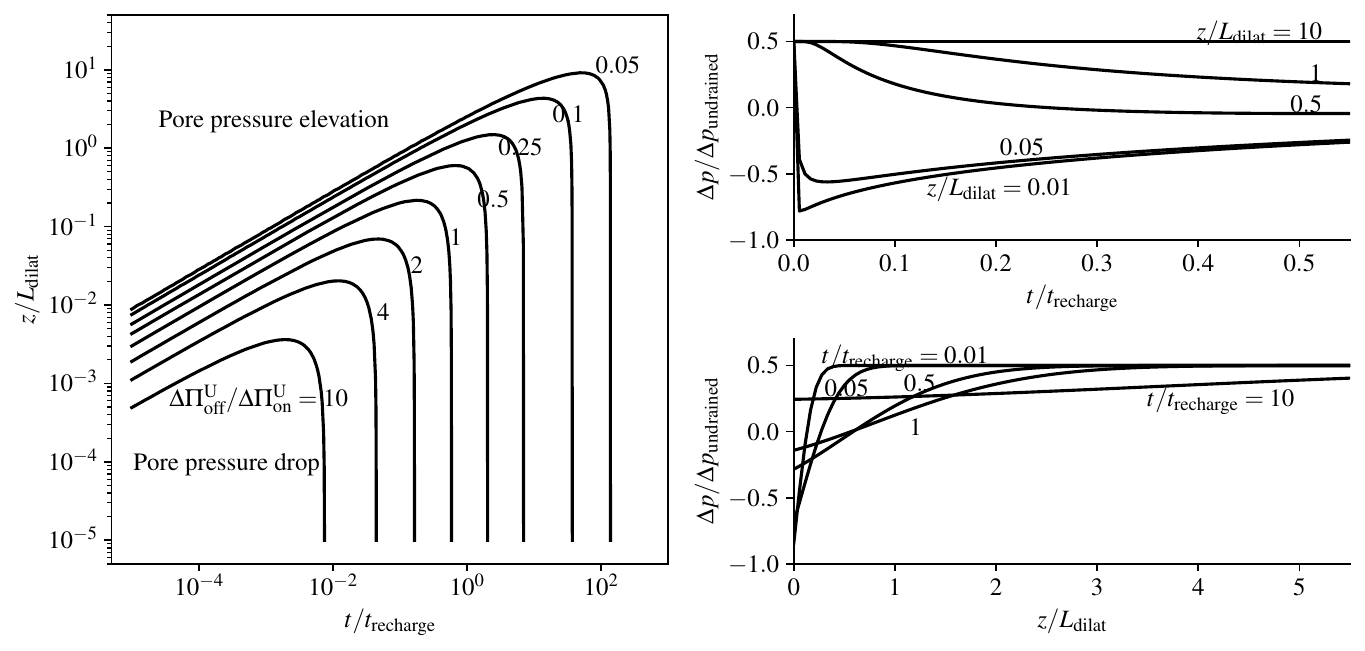}
    \caption{Evolution of the off-fault pore pressure perturbation $\Delta p/\Delta \Pi_\mathrm{on}^\mathrm{U}$ upon a step-like shear stress drop with total pore pressure drop in the fault zone ($p_f/\Delta \Pi_\mathrm{on}^\mathrm{U} \leq 1$). (a) Contours of the zero pore pressure perturbation ($\Delta p=0$) for different values of $\Delta \Pi_\mathrm{off}^\mathrm{U}/\Delta \Pi_\mathrm{on}^\mathrm{U}$. (b) Spatial and temporal evolution of the pore pressure for $\Delta \Pi_\mathrm{off}^\mathrm{U}/\Delta \Pi_\mathrm{on}^\mathrm{U}=0.5$.}
    \label{fig:parametricspace}
\end{figure}

We illustrate in Figure~\ref{fig:parametricspace} the spatiotemporal evolution of such a pressure perturbation upon a step-like shear stress drop using Eq.~\eqref{eq:InfiniteAnalytical}. We show that there exists a region near the fault zone where the pore pressure presents a negative perturbation. Such a perturbation lasts for a longer time ($t/t_\mathrm{recharge}$) and influences more volume ($z/L_\mathrm{dilat}$) with a smaller value of $\Delta \Pi_\mathrm{off}^\mathrm{U}/\Delta \Pi_\mathrm{on}^\mathrm{U}$ but tends to disappear with time due to fluid recharge from the fault walls. Beyond this region, the pore pressure experienced an elevation and the pore pressure distribution tends to the homogenized state at large time. 

\subsection{Possible impacts on the instability of a neighbouring fault}

Upon shear failure, there exists a region off the fault characterised by pore pressure elevation due to bulk porosity reduction. Such an increase in pore pressure may destabilize other faults that are more poorly oriented (with respect to the principal stress directions) than the original one and are initially locked in the first place.
In the following, we discuss such a possibility by considering purely frictional faults with zero cohesion. We assume that the faults with different orientations share the same stress field and frictional properties and that the differential stress level is sufficiently high so that the maximum stress orientations remain the same even after a local stress drop on one of the faults. 

\begin{figure}
\centering
\begin{tabular}{cc}
    \includegraphics[width=0.45\linewidth]{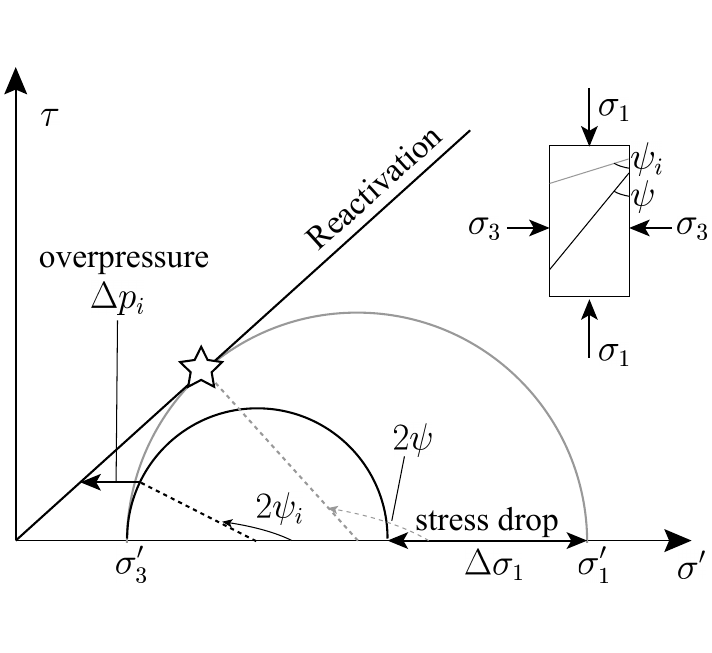} &  \includegraphics[width=0.45\linewidth]{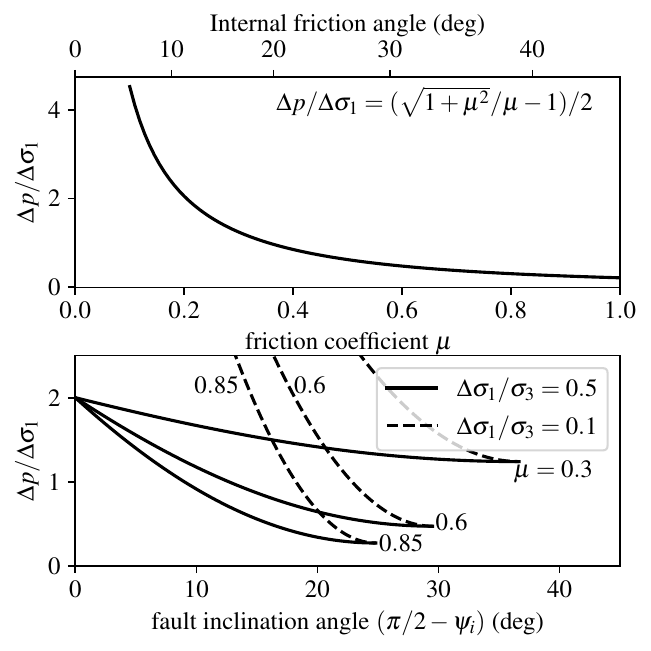} \\
    (a) & (b) 
\end{tabular}
    \caption{(a) Illustration of the pore pressure elevation necessary to destabilize a neighbouring fault which is more poorly-oriented. 
    (b) The upper panel shows the pore pressure elevation necessary to destabilize the fault with an inclination of $\pi/4-\text{arctan}(\mu)/2$ (equivalent to the inclination of a naturally formed shear band). The lower panel shows the pore pressure elevation to destabilize faults with different inclinations and friction coefficients upon stress drop ($\Delta \sigma_1/\sigma_3=0.1$ for solid curves and $\Delta \sigma_1/\sigma_3=0.5$ for dashed curves). Note that we assume here that faults are purely frictional with zero cohesion.} 
    \label{fig:Interactions}
\end{figure}

We focus on the typical case in the laboratory, where the sample experiences a drop in the axial stress with the confining stress kept constant ($\Delta \sigma_3=0, \Delta(\sigma_1-\sigma_3)=\Delta \sigma_1$) as shown in Figure~\ref{fig:Interactions}. 
We assume that the destabilized fault associated with the stress drop $\Delta \sigma_1$ shares the same inclination as that of a naturally formed shear band or ruptured fault. For purely frictional faults, the fault inclination angle ($\pi/2-\psi$) is solely determined by the friction coefficient $\mu$ with $\psi=(\mathrm{arctan}(\mu)+\pi/2)/2$ (see more details in Appendix.~\ref{ax:Interactions}).
We assume that the stress drop $\Delta \sigma_1$ is much faster than the fluid diffusion between the two faults. As a result, the pore pressure variation inside the poorly oriented fault only results from the bulk porosity reduction. We then obtain the pore pressure elevation necessary to destabilize the faults with different inclinations upon an instability-induced stress drop (Figure~\ref{fig:Interactions}). 

We first discuss the pore pressure elevation necessary $\Delta p$ to reactivate the previously destabilized fault. Assuming that there is no dilation or compaction inside the fault zone during the stress drop ($\Delta \phi_\mathrm{on}=0$), we find that the necessary pore pressure increase to reactivate the fault is merely a function of the friction coefficient (see more details in Appendix.~\ref{ax:Interactions}).
\begin{equation}
    \Delta p/\Delta \sigma_1=(\sqrt{1+\mu^2}/\mu-1)/2.
    \label{eq:interactions}
\end{equation}
A neighbouring fault that is more poorly oriented than the ruptured fault ($\psi_i>\psi$) requires a more significant pore pressure elevation than Eq.~\eqref{eq:interactions} to be destabilized (Figure~\ref{fig:Interactions}). As a result, Eq.~\eqref{eq:interactions} corresponds to the minimum requirement to possibly destabilize any neighbouring faults. 
A fault with a friction coefficient of $\mu=0.6$ requires a minimum value of $\Delta p/\Delta \sigma_1=0.47$ to be reactivated. This value becomes $\Delta p/\Delta \sigma_1=1.24$ for a fault with lower friction of $\mu=0.3$. We can thus conclude that faults presenting lower friction need larger values of $\Delta p/\Delta \sigma_1$ to trigger off-fault instability.
The reactivation of neighbouring faults also depends on the stress drop amplitude (Figure~\ref{fig:Interactions}): A larger stress drop tends to decrease the threshold $\Delta p/\Delta \sigma_1$ above which reactivation is possible. Assuming a confining stress of $\sigma_3=100$~MPa and a differential stress drop of $\Delta \sigma_1=10$~MPa, a medium value of $\Delta p/\Delta \sigma_1=0.5-1$ is sufficient to trigger the instability of a neighboring fault with an inclination angle of $23-30 \deg$ and a medium friction coefficient $\mu \approx 0.6$ (Figure~\ref{fig:Interactions}).

An off-fault coseismic pore pressure increase provides another possibility of triggering the instability of neighbouring faults and aftershocks located far away from the ruptured fault.
However, cascade fault reactivation seems unlikely: (1) The newly triggered instability presents a lower stress drop and pore pressure elevation. (2) Only a more poorly oriented fault can be destabilized, which, on the other hand, requires a more significant pore pressure elevation than that required by the previously destabilized faults. (3) Once triggered, the fault may slip and dilate making it difficult to sustain the fluid pressure elevation. (4) Moreover, the newly formed surrounding damage zone will contribute to a more drained condition by enhancing fluid diffusion. 

It is worth noticing that most experiments characterised with an off-fault pore pressure elevation present a value of $\Delta p/\Delta \sigma_1=0.11-0.14$ with a significant stress drop $\Delta \sigma_1/ \sigma_3>1$. This pore pressure increase coupled with the stress drop is insufficient to destabilise a neighbouring fault with any arbitrary inclination based on the analysis shown in Figure~\ref{fig:Interactions}. However, the off-fault pore pressure elevation $\Delta p/\Delta \sigma_1$ is probably lower in the experiments than in nature due to an imperfectly undrained system and pressure perturbations by the pore pressure sensors. We can therefore not completely rule out this off-fault pore pressure elevation as a possible mechanism to trigger the instability of a neighbouring fault originally at rest. Moreover,
such a conclusion hinges on the assumption that all faults are subjected to the same and uniform stress field. However, in nature, stress fields tend to be considerably more intricate, often exhibiting substantial heterogeneity. As a consequence of fault instability, local stress concentrations may form or relocate following a stress release. Such stress variations combined with the elevated pore pressure reported in this study may potentially contribute to the occurrence of aftershock sequences.

\subsection{Implications for earthquakes in nature}
We have demonstrated that contrasting pore pressure changes on and off the fault occurred during earthquakes in the laboratory, and we have reproduced a similar evolution of the pore pressure using a simplified analytical model.  However, laboratory observations and model predictions do not guarantee the occurrence of the same phenomena during natural earthquakes. Whether the off-fault pore pressure elevation could manifest itself during earthquakes or instability in nature still remains unanswered.

To address this question, we first examine the two essential conditions required for the presence of off-fault pore pressure elevation, and then discuss the observations which could potentially capture the reported mechanism during earthquakes.

\subsubsection{Examination of the conditions for off-fault pore pressure elevation}
The increase in off-fault pore pressure necessitates the porosity reduction from a relatively high stress level and undrained conditions during shear failure or instability. Here, we aim to determine the specific circumstances where such a phenomenon would favourably occur in nature.

The stress state of a pre-existing natural fault at rest is constrained by its sliding resistance, setting an upper limit on the potential fault stress level. As previously discussed, this upper limit aligns closely with the onset of dilatancy \citep{Hadl1973}, especially when the fault exhibits an orientation akin to that of a naturally formed shear failure band. Consequently, the stress levels within most natural faults tend to approximate or fall below the onset of dilatancy, unless they are not optimally oriented for slip. Additionally, the maximum stress drop associated with natural earthquakes typically ranges in the order of a few to tens of MPa \citep[see][and references therein]{StCo2001}. It is thus uncommon for natural faults to exhibit the same combination of a substantial stress drop and high stress level as observed in laboratory experiments. However, exceptions may arise in cases of geometrical irregularities, such as rough, bent, or kinked faults. These irregularities can lead to stress levels surpassing the threshold for dilatancy initiation, increasing the likelihood of observing the mechanism proposed in this study, particularly in localised regions.

The off-fault pore pressure elevation also necessitates porosity reduction and undrained conditions during shear failure. Not all types of rocks can accommodate these conditions. For most compact brittle polycrystalline materials, dilatancy associated with anisotropic microcracking prior to macroscopic failure in compression appears to be a general feature \citep[][Chapter 5 and references therein]{PaWo2005}. Such dilatancy also occurs in porous rocks, such as sandstone. However, the increase in pore volume may be counteracted by inelastic pore collapse during deformation. Furthermore, the relatively higher porosity/permeability in these porous rocks \citep[][Chapter 5]{PaWo2005} make it more likely to exhibit drained conditions, characterised by a significantly smaller diffusion time scale. As a result, compact rocks such as granite and carbonate, compared with sandstone, are more likely to experience an elevation in off-fault pore pressure upon stress drop and sustain this increase over time.

\subsubsection{Possible observations to capture the reported mechanism in nature} 
Although a few documented cases of aftershocks induced by pore pressure increase exist in earthquakes with complex rupture geometries \citep{HaUp2018,ShWe2020,TaAl2023},  there is currently no evidence showing that the reported elevated pore pressure resulted from the closure of the dilatant microcracks associated with the mainshock stress drop. To our knowledge, the direct observation in nature supporting the mechanism reported in this study are missing. Here, we highlight some observational approaches that might reflect off-fault pore pressure elevation in natural settings. 

The spatio-temporal evolution of shallow pore pressure variations during earthquake cycles is commonly tracked through changes in groundwater levels within a network of wells \citep{Wang2001,Chia2008}. 
Despite the limited depth of these groundwater wells, typically extending only a few hundred meters, they provide insights into pore pressure variations near a hypocenter through fluid diffusion, reflecting processes like microcrack opening or closure. However, these groundwater level changes may also be sensitive to shaking-induced diffusivity increases or porosity and permeability changes in near-surface sediments or aqueducts \citep{WaWa2004,WaMa2021}. Particular caution is thus needed to rule out possibilities related to changes in near-surface properties. Satellite radar interferograms (InSAR) \citep{ZeRo1994} offer an alternate method to assess the pore pressure variation field by measuring surface displacements \citep{PeRR1998,JoSe2003} induced by groundwater flow and time-dependent strain \citep{Roel1996}. Similar to the groundwater level monitoring, this method necessitates additional efforts to account for drainage effects \citep{JoSe2003} and changes in properties of shallow rocks.

Seismic velocities provide a valuable avenue to unveil the temporal evolution of dilatant microcracks \citep{BoZo2004,Elsi2023} and pore pressure \citep{ToSi1972,Chri1984,DaRe2000}. Depending on the dominant microcrack orientation, damaged rock may become anisotropic, and shear wave splitting might occur \citep{Cram1987,Cram1994}, which could be linked to orientation and magnitude of differential stress \citep{ZaCr1997}. As a result, seismological analysis has the potential to capture the off-fault pore pressure elevation during seismic activities by examining microcrack orientations and their associated stress levels. In general, a comprehensive approach that combines multiple geophysical datasets (geodetic, hydrological, seismological) would be required to detect in the field the effect observed in our laboratory experiment. In practice, the main limitation is the space and time scales over which the pore pressure rise occurs, which could be too small to have a direct detectable signature.

\section{Conclusions}

We conducted laboratory rupture experiments  under triaxial conditions to investigate the evolution of pore pressure heterogeneity around faults during slip. Our findings reveal notable pore pressure perturbations within the fault zone and the adjacent bulk material, exhibiting opposite trends. This results in significant gradients in pore pressure occurring over small distances.

The observed drop in pore pressure along the fault primarily stems from a porosity increase associated with fracture growth, fracture coalescence, and fault dilation, consistent with previous measurements \citep{Bran2020,AbBr2021,AbBr2023}. Conversely, the rise in pore pressure away from the fault is attributed to a reduction in bulk porosity, a consequence of the closure of previously formed dilatant microcracks, which predominantly align parallel to the maximum compression axis.

Using a simplified analytical model, we demonstrate that off-fault pore pressure increase during rupture necessitates relatively undrained conditions of fluid flow and a concurrent reduction in bulk porosity during shear failure or fault instability. Achieving the former condition is feasible in nature, particularly in scenarios involving earthquakes propagating in low-permeability wall rocks. However, satisfying the latter condition necessitates a high stress level surpassing the onset of dilatancy. While such elevated stress levels may be uncommon in the context of most natural faults, they may manifest locally, owing to the inherent irregularities in fault geometry.

Our measurements serve as an empirical foundation for understanding the marked heterogeneity in pore pressure dynamics during seismic events, where the fault acts as a localised sink for pore fluids, in contrast to the surrounding bulk material which acts as a dispersed source. This pronounced heterogeneity in pore pressure distribution may have a seismic signature through its effects on seismic wave velocities. Additionally, it could contribute to the occurrence of afterslip \citep[e.g.][]{AbBr2023} and/or aftershock sequences when coupled with stress field perturbations.

\section*{Acknowledgement}
Funding by the European Research Council under the European Union's Horizon 2020 research and innovation programme (project RockDEaF, grant agreement \#804685) and the UK Natural Environment Research Council (grant NE/S000852/1) is acknowledged. We also thank Christopher Harbord, Thomas M. Mitchell, and Bobby Elsigood for their help in performing the experiments, and Dmitry Garagash for insightful discussions.

\section*{Data availability}
The experimental data shown in this paper is available on the Zenodo platform with the identifier\\~\url{https://doi.org/10.5281/zenodo.10553131}.

\section*{Author contributions}
DL: Conceptualization, methodology, software, validation, formal analysis, investigation, visualization, writing—original draft. NB: Conceptualization, methodology, investigation, supervision, resources, writing—review and editing. FA: Conceptualization, software, writing—review and editing.

\appendix

\section{Fault geometry and location of pore pressure transducers}\label{ax:geometrylocations}
WG17 and WG18 presented planar fault planes while WG16 and WG21 showed more geometrical irregularities, see Figure~\ref{fig:faultgeometry}. Veins were found in WG16 and WG17. They were located in a different region than the fault and did not seem to influence the position of the fault planes. When the fault geometry is complex, we approximate it as a planar fault plane with an angle of $\psi$ with respect to the maximum compression axis. The distances between the different pore pressure sensors and the fault plane can be calculated and are shown in Table.~\ref{tab:faultgeometry}.

\begin{table}[!htp]
    \centering
    \begin{tabular}{c|cccc}
    \hline
         Sample & WG16 & WG17 & WG18 & WG21\\
    \hline
    $\psi$ (deg) & 30 & 30 & 27 & 27.5\\
    Transducer (1) & $\approx$ 0 & 0 & 0 & $\approx$ 0 \\
    Transducer (2) & 0 & 0 & 0 & 19.06 \\
    Transducer (3) & 0 & 0.89 & 6.93 &  $\lessapprox$ 28.63\\
    Transducer (4) & 0 & 4.82 & 10.86 & $\approx$ 16.30\\
    Transducer (5) & 1.70* & 15.36 & 16.60 &  - \\
    Transducer (6) & 7.65 & - & 22.30 & - \\
    \hline
    \end{tabular}
    \caption{Distances (in mm) of the pore pressure sensors to the approximated planar fault. The locations of the numbered transducers are illustrated in  Figure~\ref{fig:faultgeometry}. When the pore pressure sensors are in direct hydraulic connection with the fault via fractures, we consider that their distance to the fault plane is zero. * indicates the distance between the transducer to the nearest horizontal fracture which is hydraulically connected to the fault. 
    }
    \label{tab:faultgeometry}
\end{table}

 The shear stress $\tau$ and normal stress $\sigma_n$ of the inclined fault can be approximated as follows.
\begin{equation}
    \begin{aligned}
        &\tau= \frac{q}{2}\sin (2\psi),\\
        &\sigma_n=P_c+\frac{q}{2}(1-\cos (2\psi)).
    \end{aligned}
    \label{eq:Triaxstress}
\end{equation}
The slip $\delta$ along the fault is corrected with the elastic deformation of the loading column.
\begin{equation}
    \delta= (s - q A/k)/\cos(\psi),
\end{equation}
where $k=480$~kN/mm is the stiffness of the loading column, $s$ is the axial shortening measured by external LVDTs, and $A$ is the section area of the cylindrical sample.

\section{Estimation of the bulk porosity change}\label{ax:porosity}
We approximate the porosity change upon shear stress drop using the volumetric strain variation measured from cyclic loading tests in dry Westerly granite.
The bulk porosity change following the main failure can be estimated using samples that experienced a maximum differential stress very close to the peak failure stress. Some of the bulk porosity changes can be reversible, and some are not. In this paper, we assume that the porosity change is a combination of reversible and irreversible changes \citep{StLo2003,BrPe2023} and approximate it using the volumetric strain evolution during the unloading from the maximum differential stress in each cycle of the loading tests \citep{BrPe2023}.
\begin{equation}
 \phi(q)-\phi(q=0)\approx \epsilon_v^t- \epsilon_v^{res}-\frac{q}{3K_m},
\end{equation}
where $\epsilon_v^{t}$ is the volumetric strain at the peak stress containing the contribution of elastic crack opening, inelastic slip and crack growth, and the elastic volumetric deformation of the solid matrix. $\epsilon_v^{t}$ can be obtained from the volumetric strain measured during the unloading phase as shown in figure~12 of \cite{BrPe2023}. $\epsilon_v^{res}$ indicates the residual volumetric strain after unloading, which is reported in the unloading branch of the data in figure~3 of \cite{BrPe2023}. The last term $q/(3K_m)$ represents the elastic volumetric strain induced by the matrix deformation, where $q$ is the differential stress and $K_m$ the bulk elastic modulus of the solid matrix ($K_m=45$~GPa for Westerly granite \citep{JaCZ2009}). 

\begin{figure}[htp]
\centering
    \includegraphics[width=0.45\linewidth]{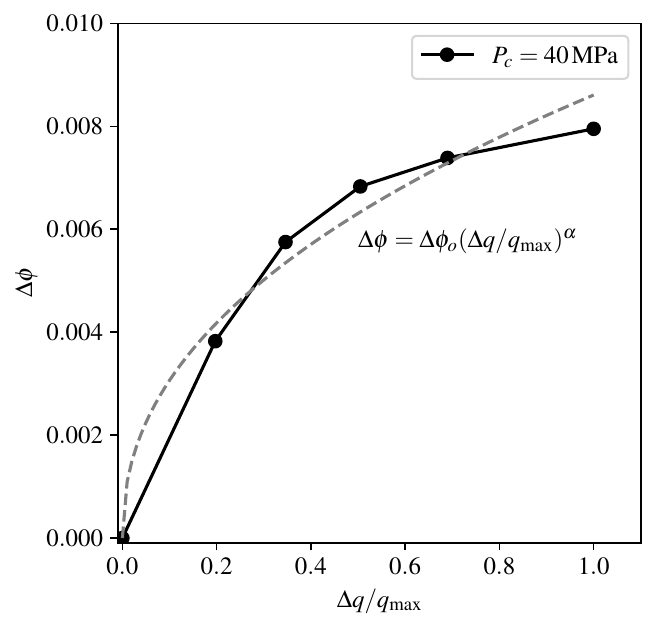} 
    \caption{Estimated porosity change during unloading for dry Westerly granite under an effective confining stress of 40~MPa \citep{BrPe2023}. The gray dashed curves correspond to the expression of Eq.~\eqref{eq:inelasticporosity} with $\Delta \phi_o=0.0086$ and $\alpha=0.449$.}
    \label{fig:poroinelasticityfitting}
\end{figure}

In the following, we define the porosity change inside the bulk $\Delta \phi$ as the change of the porosity with respect to its value $\phi^\mathrm{ref}(q_\mathrm{max})$ at the peak differential stress $q_\mathrm{max}$. 
\begin{equation}
    \Delta \phi=\phi^\mathrm{ref}(q_\mathrm{max})-\phi(q).
\end{equation}
We plot in Figure~\ref{fig:poroinelasticityfitting} the estimated porosity change as a function of the differential stress in dry Westerly granite under a constant confining stress of 40~MPa, which corresponds to the same effective confining pressure in this study. The bulk porosity change can be approximated as a power-law function of the differential stress drop.
\begin{equation}
    \Delta \phi=\Delta \phi_o(\Delta q/q_\mathrm{max})^\alpha,
    \label{eq:inelasticporosity0}
\end{equation}
where $\Delta \phi_o$ represents the porosity variation upon complete unloading ($\Delta q/q_\mathrm{max}=1$) which depends on the material and effective confining stress. $\alpha$ is a material-dependent parameter to describe the power-law-like dependence of porosity changes on the differential stress. For Westerly granite under an effective confining stress of 40~MPa, we obtain $\Delta \phi_o=0.0086$ and $\alpha=0.449$ by fitting the experimental data (Figure~\ref{fig:poroinelasticityfitting}).
We then obtain the porosity variation in presence of pore fluids. 
\begin{equation}
    \Delta \phi=\Delta \phi_\mathrm{off}^\mathrm{mid} (1-\mathrm{e}^{(p-P_c)/P_o})(\Delta q/q_\mathrm{max})^\alpha, \, \Delta \phi_\mathrm{off}^\mathrm{mid}=\frac{\Delta \phi_o}{1-\mathrm{e}^{(p_f-P_c)/P_o}},
    \label{eq:inelasticporosity1}
\end{equation}
where $p$ is the local pore pressure and $p_f$ is the constant pore pressure applied at the sample ends. $P_o$ is a pressure scale dependent on the material and confining stress describing the sensitivity of porosity variation to the change of effective confining stress (In the case of a constant confining stress, it characterises the corresponding sensitivity to the change of local pore pressure). When $P_o>0$, a decrease in local pore pressure favours the porosity reduction inside the bulk, and an increase in local pore pressure prohibits the porosity reduction. 

The porosity variation \eqref{eq:inelasticporosity1} reduces to the same expression as in \eqref{eq:inelasticporosity0} for dry Westerly granite $p=p_f=0$ and uniformly saturated Westerly granite $p=p_f$. Assuming that the pore pressure stays below the confining stress ($p\in (0, P_c)$), we obtain the possible range of the maximum porosity reduction of $(0, \Delta \phi_o (1-\mathrm{e}^{-P_c/P_o})/(1-\mathrm{e}^{(p_f-P_c)/P_o}))$. 

\section{Numerical discretization \label{ax:numerical}}
\subsection{Finite sample size}
\paragraph{Transient porosity evolution}
We obtain the dimensionless form of the governing equations for the fluid flow.
\begin{align}
    \frac{\partial p}{\partial t}- \frac{\partial^2 p}{\partial z^2}=\mathcal{G}_p \Delta \dot{\phi}, \, \mathrm{for}\, z>0, \label{eq:gveq1}\\
     \frac{\partial p|_{z=0}}{\partial t}-\mathcal{G}_l\left.\frac{\partial p}{\partial z}\right|_{z=0}=-h^{\prime}\left(t \mathcal{G}_t \right) \mathcal{G}_t, \label{eq:gveq2}\\
    p(1,t)=p(z,0)=\mathcal{G}_f, \label{eq:gveq3}
\end{align}
\begin{equation}
\begin{aligned}
\Delta \dot{\phi}=(\mathcal{G}_s f(t\mathcal{G}_t))^\alpha \left(-\frac{\dot{p}}{\mathcal{G}_o} \mathrm{e}^{(p-\mathcal{G}_c)/\mathcal{G}_o}+\alpha \frac{\mathcal{G}_t f^\prime(t\mathcal{G}_t)}{f(t \mathcal{G}_t)} (1-\mathrm{e}^{(p-\mathcal{G}_c)/\mathcal{G}_o})\right), \label{eq:gveq4}
\end{aligned}
\end{equation}
with
\begin{equation}
    \mathcal{G}_p=\frac{\Delta \Pi_\mathrm{off}^\mathrm{U}}{\Delta \Pi_\mathrm{on}^\mathrm{U}},\, 
    \mathcal{G}_l=\frac{2 S L}{S_f w},\,  \mathcal{G}_t=\frac{t_\mathrm{diff}}{t_\mathrm{fail}}, \, 
    \mathcal{G}_f=\frac{p_f}{\Delta \Pi_\mathrm{on}^\mathrm{U}}, \,  
    \mathcal{G}_c=\frac{P_c}{\Delta \Pi_\mathrm{on}^\mathrm{U}}, \, \mathcal{G}_s=\frac{\Delta q_\mathrm{max}}{q_\mathrm{max}}, \, \mathcal{G}_o=\frac{P_o}{\Delta \Pi_\mathrm{on}^\mathrm{U}}.
    \label{eq:DmsGroups2}
\end{equation}
We compute the pressure rate in the fault by solving the full diffusion problem using finite differences in space and the method of lines following the same procedure as in \cite{AbBr2021}.
We discretize space into $N+1$ nodes at positions $z_0,\ldots,z_N$ uniformly spaced with spacing $\Delta z$, and use centered finite differences to evaluate spatial derivatives. $z_0$ represents the fault position and $z_N$ the sample end. For all numerical results shown in this study, we use $N=501$.
We obtain in the following the discretized form of the diffusion equation accounting for a transient variation of porosity.
\begin{equation}\label{eq:FD}
\begin{aligned}
  \frac{dp_n}{dt} =\left(\frac{p_{n+1} - 2p_n + p_{n-1}}{\Delta z^2}+\alpha \mathcal{G}_p (\mathcal{G}_s f(t \mathcal{G}_t))^\alpha \frac{\mathcal{G}_t f^{\prime}(t \mathcal{G}_t)}{f(t\mathcal{G}_t)}(1-\mathrm{e}^{(p_n-\mathcal{G}_c)/\mathcal{G}_o})\right)\\
  /\left(1+ \frac{\mathcal{G}_p(\mathcal{G}_s f(t \mathcal{G}_t))^\alpha}{\mathcal{G}_o} \mathrm{e}^{(p_n-\mathcal{G}_c)/\mathcal{G}_o} \right),
\end{aligned}
\end{equation}
The boundary conditions read
\begin{equation}\label{eq:BCtop_FD}
  \frac{dp_N}{dt} = 0,
\end{equation}
\begin{equation}\label{eq:BCbottom_FD}
\begin{aligned}
  \frac{\text{d}p_0}{\text{d}t} = \left[ \frac{p_1-p_0}{\Delta z}-\frac{1}{\mathcal{G}_l}h'(t \mathcal{G}_t)\mathcal{G}_t \right]/\left(\frac{1}{\mathcal{G}_l}+\frac{1}{2}\Delta z \right).
\end{aligned}
\end{equation}

Equations \eqref{eq:FD}, \eqref{eq:BCtop_FD} and \eqref{eq:BCbottom_FD} form a system of $N+1$ ODEs in time for unknowns $\{p_0,\ldots,p_N\}$. 
We solve this system using a 5th order, A-L stable, stiffly-accurate, explicit singly diagonal implicit Runge-Kutta method with splitting (see \cite{kennedy03}, method implemented as \verb+KenCarp5()+ in the \verb+DifferentialEquations.jl+ package, see \cite{rackauckas17}).

\paragraph{Instantaneous porosity evolution}
In case of a dynamic failure $t_\mathrm{fail} \to 0$ and an instantaneous porosity change, we simulate the step-wise evolution of the shear stress drop and the porosity variation by changing the initial conditions. The governing equations can be simplified as follows.  
\begin{align}
    & \frac{\partial p}{\partial t}-\frac{\kappa}{\eta S} \frac{\partial^2 p}{\partial z^2}=0,\\
    & \frac{\partial p|_{z=0}}{\partial t}-\frac{2 \kappa}{S_f w \eta}\left.\frac{\partial p}{\partial z}\right|_{z=0}=0,\\
    & p(z=L,t)=p_f.\\
    \label{eq:fluidmass1}
\end{align}
Initial conditions read 
\begin{align}
    & p(z=0, 0^+)=p(z=0, 0^-)-\Delta p_\mathrm{on}^\mathrm{U},\\
    & p(0<z<L, 0^+)=p(0<z<L, 0^-)+\Delta p_\mathrm{off}^\mathrm{U},
    \label{eq:faultfluidmass1}
\end{align}
where $\Delta p_\mathrm{on}^\mathrm{U}$ and $\Delta p_\mathrm{off}^\mathrm{U}$ are the coseismic pore pressure increase and decrease. 
Following Eq.~\eqref{eq:scalequantities} but scaling the pressure using $\Delta p_\mathrm{on}^\mathrm{U}$, we obtain the discretized form of the governing equations.
\begin{align}
      &\frac{dp_n}{dt} = \frac{p_{n+1} - 2p_n + p_{n-1}}{\Delta z^2}, \, 0<n<N,\\
      &\frac{\text{d}p_0}{\text{d}t} = \left[ \frac{p_1-p_0}{\Delta z}\right]/\left(\frac{1}{\mathcal{G}_l}+\frac{1}{2}\Delta z \right), \\
      &\frac{dp_N}{dt} = 0,
\end{align}
and the initial conditions become dimensionless with the pressure scaled by $\Delta p_\mathrm{on}^\mathrm{U}$.
We then adopt the same solving procedure as the case of a transient porosity change.

\subsection{Infinite medium}
We present in the following the formulation and its numerical discretization for the case of an infinite medium accounting for a transient porosity change. We map the infinite medium onto the range of $(0,1)$ following 
\begin{equation}
    \xi \leftarrow 1/(z/L_\mathrm{dilat}+1), \,  t \leftarrow t/t_\mathrm{recharge}, \, p \leftarrow p/\Delta \Pi_\mathrm{on}^\mathrm{U}, \, \Delta \phi \leftarrow \Delta \phi_\mathrm{off}/\Delta \phi_\mathrm{off}^\mathrm{mid}.
    \label{eq:scalequantitiesInf}
\end{equation}
We obtain the dimensionless form of governing equations.
\begin{align}
    \frac{\partial p}{\partial t}- \left(\xi^4 \frac{\partial^2 p}{\partial \xi^2}  +2 \xi^3 \frac{\partial p}{\partial \xi} \right)=\mathcal{G}_p \Delta \dot{\phi}, \, \mathrm{for}\, \xi<1,\\
     \frac{\partial p|_{\xi=1}}{\partial t}+\xi^2\left.\frac{\partial p}{\partial \xi}\right|_{\xi=1}=-h^{\prime}\left(t \mathcal{G}_r \right) \mathcal{G}_r,\\
    p(0,t)=p(\xi,0)=\mathcal{G}_f,
\end{align}
\begin{equation}
\begin{aligned}
\Delta \dot{\phi}=(\mathcal{G}_s f(t\mathcal{G}_r))^\alpha \left(-\frac{\dot{p}}{\mathcal{G}_o} \mathrm{e}^{(p-\mathcal{G}_c)/\mathcal{G}_o}+\alpha \frac{\mathcal{G}_r f^\prime(t\mathcal{G}_r)}{f(t \mathcal{G}_r)} (1-\mathrm{e}^{(p-\mathcal{G}_c)/\mathcal{G}_o})\right),
\end{aligned}
\end{equation}
where $\mathcal{G}_r=t_\mathrm{recharge}/t_\mathrm{fail}$. The other dimensionless groups are the same as those defined in Eq.~\eqref{eq:DmsGroups2}. 
We discretize the domain into $N+1$ nodes at positions $\xi_0,\ldots,\xi_N$ with spacing $\Delta \xi$. We use centered finite differences to evaluate spatial derivatives. $\xi_0$ represents the far field location and $\xi_N$ the fault position. 

\begin{equation}
\begin{aligned}
  \frac{dp_n}{dt} =\left(((n+1)p_{n+1} - 2n p_n + (n-1)p_{n-1})\Delta \xi^2 n^3+\alpha \mathcal{G}_p (\mathcal{G}_s f(t \mathcal{G}_t))^\alpha \frac{\mathcal{G}_t f^{\prime}(t \mathcal{G}_t)}{f(t\mathcal{G}_t)}(1-\mathrm{e}^{(p_n-\mathcal{G}_c)/\mathcal{G}_o})\right)\\
  /\left(1+\frac{\mathcal{G}_p(\mathcal{G}_s f(t \mathcal{G}_t))^\alpha}{\mathcal{G}_o} \mathrm{e}^{(p_n-\mathcal{G}_c)/\mathcal{G}_o}\right),
\end{aligned}
\end{equation}
\begin{equation}
  \frac{dp_0}{dt} = 0,
\end{equation}
\begin{equation}
\begin{aligned}
  \frac{\text{d}p_N}{\text{d}t} = \left[  N^4\frac{p_{N-1}-p_N}{1/\Delta \xi}-N(N+1)h'(t \mathcal{G}_r)\mathcal{G}_r  \right]/\left(\frac{1}{2}\frac{1}{\Delta \xi}+N(N+1) \right).
\end{aligned}
\end{equation}
We adopt the same solving procedure as the case of a finite-size sample.

\section{Parametric study of dimensionless groups and effect of the sample size}\label{ax:boundary}

\begin{figure}[htp]
\begin{tabular}{cc}
\centering
    \includegraphics[width=0.45\textwidth]{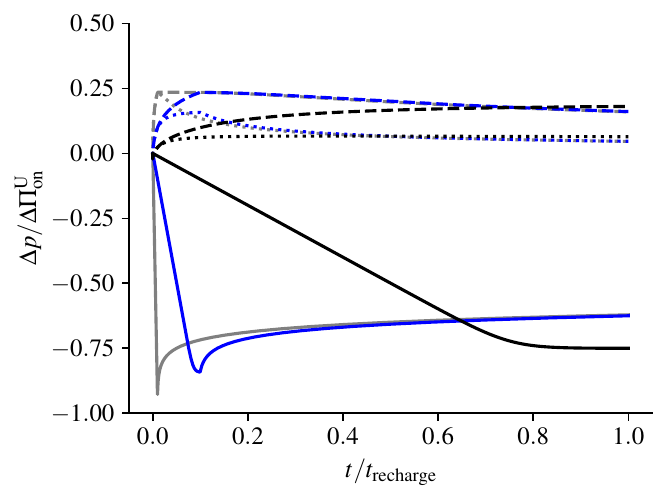}& 
    \includegraphics[width=0.45\textwidth]{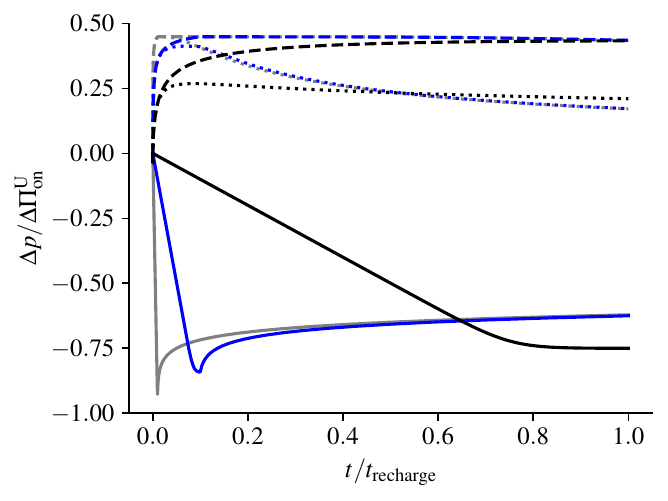}\\
    (a) Default case & (b) lower pore-pressure sensitivity $P_o/\Delta \Pi_\mathrm{on}^\mathrm{U}=0.5$ \\
    \includegraphics[width=0.45\textwidth]{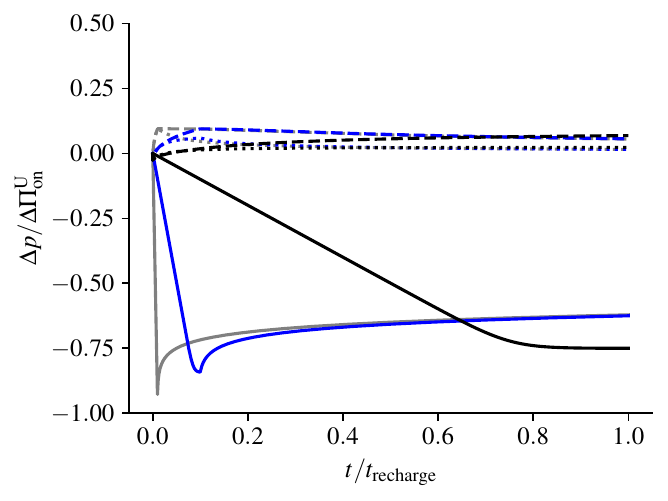}&
    \includegraphics[width=0.45\textwidth]{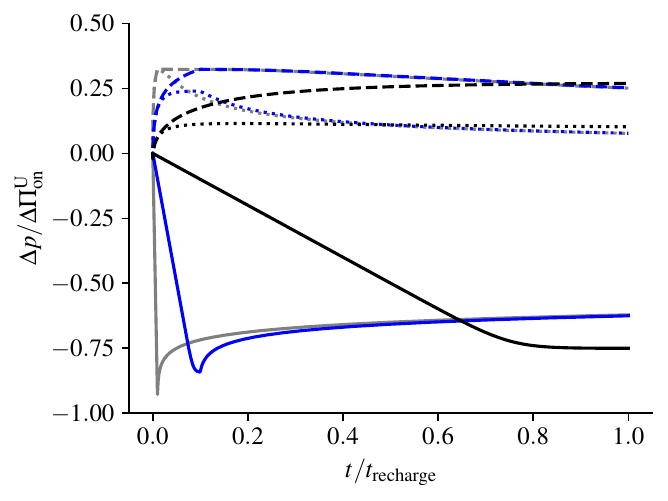}\\
    (c) smaller stress drop $\Delta q_\mathrm{max}/q_\mathrm{max}=0.1$  & (d) larger porosity reduction $\Delta \Pi_\mathrm{off}^\mathrm{U}/\Delta \Pi_\mathrm{on}^\mathrm{U}=10$
\end{tabular}
    \caption{Spatiotemporal evolution of the pore pressure at $z/L_\mathrm{dilat}=0, 0.25, 1.0$ (indicated respectively by solid, dotted and dashed curves) in an infinite medium. The gray, blue and black curves indicate different shear failure periods of $t_\mathrm{recharge}/t_\mathrm{fail}=100, 10, 1$. (a) By default, we set $p_f/\Delta \Pi_\mathrm{on}^\mathrm{U}=1.0$,  $P_c/\Delta \Pi_\mathrm{on}^\mathrm{U}=1.5$, $P_o/\Delta \Pi_\mathrm{on}^\mathrm{U}=5.0$, $\Delta q_\mathrm{max}/q_\mathrm{max}=1.0$, $\Delta \Pi_\mathrm{off}^\mathrm{U}/\Delta \Pi_\mathrm{on}^\mathrm{U}=5.0$, $\alpha=0.5$. We assume a linear evolution of the stress drop and the dilatancy-related porosity and a uniform pore pressure distribution before failure. (b) Effects of the sensitivity of the porosity variation to the change of pore pressure, (c) the stress drop, and (d) potential porosity reduction. }
    \label{fig:InfiniteEffects}
\end{figure}

For an infinite medium, we assume that the stress drop and fault dilatancy evolves linearly with time during the failure. 
We then use the 1-D fluid flow model to investigate the effects of the shear failure time, stress drop amplitude, porosity reduction quantity, and sensitivity of the porosity variation on local pore pressure.
When accounting for a transient porosity reduction as in Eq.~\eqref{eq:porosityrate}, the governing dimensionless groups remain the same as those shown in Eq.~\eqref{eq:DmsGroups}, except that $2SL/S_f w$ and $t_\mathrm{diff}/t_\mathrm{fail}$ are now replaced by one dimensionless group $t_\mathrm{recharge}/t_\mathrm{fail}$, indicating the ratio between the fluid recharge timescale and the shear failure duration.
We run a series of simulations using different dimensionless groups and illustrate their effects in Figure~\ref{fig:InfiniteEffects}. We show that a larger porosity reduction $\Delta \Pi_\mathrm{off}^\mathrm{U}/\Delta \Pi_\mathrm{on}^\mathrm{U}$, a larger stress drop $\Delta q_\mathrm{max}/q_\mathrm{max}$, a lower sensitivity of the porosity variation on local pore pressure $P_o/\Delta \Pi_\mathrm{on}^\mathrm{U}$ and a shorter failure period $t_\mathrm{recharge}/t_\mathrm{fail}$ tends to sustain a more significant increase of the pore pressure for a longer duration.  Different from the prediction of Eq.~\eqref{eq:InfiniteAnalytical}, the equilibrium state at large time now corresponds to a uniform pore pressure distribution with the same value of $p_f$ as that imposed in the far field.

\begin{figure}[htp]
    \centering
        \includegraphics[width=0.9\textwidth]{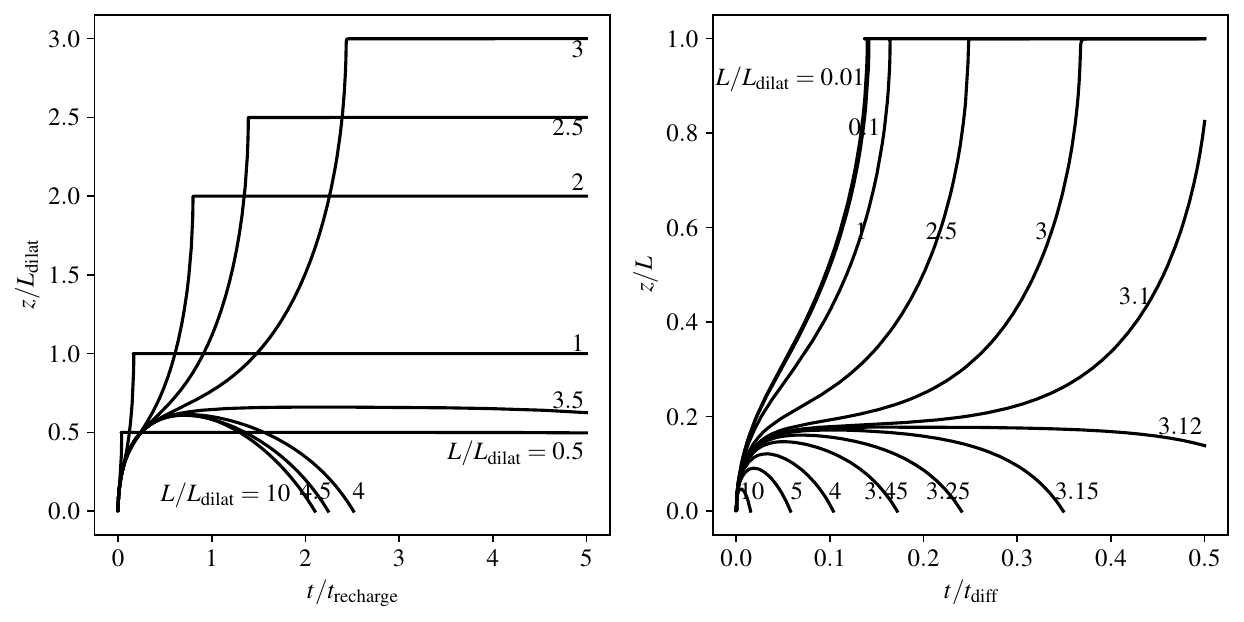} 
    \caption{Contours of the zero pore pressure perturbation ($\Delta p=0$) for $\Delta \Pi_\mathrm{off}^\mathrm{U}/\Delta \Pi_\mathrm{on}^\mathrm{U}=$0.5, $p_f/\Delta \Pi_\mathrm{on}^\mathrm{U}$=1.0  with different material properties and sample sizes $L/L_\mathrm{dilat}$. The parametric space below the contours is characterised by a pore pressure drop.
    }
    \label{fig:DiffusionIllustration}
\end{figure}

For a finite-size sample in the laboratory, the effect of the sample size $2L$ can not be disregarded when $L/L_\mathrm{dilat}\ll 1$. We illustrate this boundary effect in Figure~\ref{fig:DiffusionIllustration} for the case of $\Delta \Pi_\mathrm{off}^\mathrm{U}/\Delta \Pi_\mathrm{on}^\mathrm{U}=0.5$ considering various values of $\mathcal{G}_l=2SL/S_f w=L/L_\mathrm{dilat}$. As shown in the figure, the boundary effect becomes negligible when $L/L_\mathrm{dilat}>5$. In this scenario, the near-fault region exhibits an initial drop in pore pressure, succeeded by an increase attributed to bulk compaction, and finally a gradual return to the equilibrium state with uniform pressure distribution. Conversely, the off-fault region experiences an initial rise in pore pressure, followed by a gradual return to equilibrium. 
When the sample size comes into play ($L/L_\mathrm{dilat}<5$), the near-fault region initially experiences a pore pressure drop after the failure, which is then followed by a gradual increase back to the equilibrium state. 
The off-fault region first sustains elevated pore pressure for a while and eventually transitions to the equilibrium state at large time.
This sustained elevation is maintained for a duration of at least $0.15 t_\mathrm{diff}=0.15 \eta S L^2/\kappa$ as shown in Figure~\ref{fig:DiffusionIllustration}. 
This suggests that the maximum response time of the pore pressure sensors should not go beyond $0.15 t_\mathrm{diff}$ in order to detect the off-fault pore pressure elevation in small-size laboratory samples.

\section{Mohr-Coulomb fault instability analysis}\label{ax:Interactions}
We assume that the faults are purely frictional with zero cohesion and share the same friction coefficient $\mu=\tan{\beta}$. The stresses on the fault satisfy the following relation at the onset of instability.
\begin{equation}
    \frac{\sigma_1^\prime}{\sigma_3^\prime}=\frac{1+\sin{\beta}}{1-\sin{\beta}},
    \label{eq:MohrCircleFailure}
\end{equation}
where $\sigma_1^\prime$ and $\sigma_3^\prime$ are effective principal stresses (Figure~\ref{fig:Interactions}). The inclination angle of the fault with respect to the direction of the minimum confining stress thus corresponds to $\pi/4-\beta/2$. We assume that there is a differential stress drop of $\Delta \sigma_1$ ($\Delta \sigma_3=0$) due to fault instability and it will not change the direction of the maximum compression $\Delta \sigma_1 \leq 2 \sigma_3^\prime \sin{\beta}/(1-\sin{\beta})$. We obtain the pressure increase $\Delta p$ necessary to destabilize a neighbouring fault having an inclination angle of $\psi_i$ ($\psi_i<\psi$, see Figure~\ref{fig:Interactions}). 
\begin{equation}
    \Delta p (\psi_i)= \frac{\sigma_1^{\prime}-\Delta\sigma_1+\sigma_3^{\prime}}{2}-\frac{\sigma_1^{\prime}-\Delta\sigma_1-\sigma_3^{\prime}}{2} \cos{2(\frac{\pi}{2}-\psi_i)}-\frac{\sigma_1^{\prime}-\Delta\sigma_1-\sigma_3^{\prime}}{2}\sin{2(\frac{\pi}{2}-\psi_i)}\cot{\beta}.
    \label{eq:overpressure}
\end{equation}
Substituting Eq.~\eqref{eq:MohrCircleFailure} into Eq.~\eqref{eq:overpressure}, one obtains 
\begin{equation}
\begin{aligned}
    \frac{\Delta p (\psi_i)}{\Delta \sigma_1}=\frac{1}{\Delta \sigma_1/\sigma_3} \frac{1-\cos{\beta}\sin{2(\pi/2-\psi_i)}-\cos{2(\pi/2-\psi_i)}\sin{\beta}}{1-\sin{\beta}}\\
    +\frac{\cot{\beta}\sin{2(\pi/2-\psi_i)}+\cos{2(\pi/2-\psi_i)}-1}{2}.
\end{aligned}
\end{equation}
By setting $\psi_i=\pi/4+\beta/2$, we obtain the corresponding pore pressure elevation necessary to destabilize the fault with the same inclination as that of a naturally formed shear band.
\begin{equation}
    \frac{\Delta p (\pi/4+\beta/2)}{\Delta \sigma_1}=\frac{1}{2}\left(\frac{\sqrt{\mu^2+1}}{\mu}-1\right).
\end{equation}

\printbibliography

\end{document}